\documentclass[10pt,english,aps,prx,twocolumn,superscriptaddress]{revtex4-1}
\usepackage{amsmath}
\usepackage{color}
\usepackage{graphicx}
\usepackage{amssymb}
\usepackage{dsfont}

\def\be{\begin{equation}}
\def\ba#1{\begin{array}{#1}}
\def\ea{\end{array}}
\def\ee{\end{equation}}
\def\l{\left}
\def\rr{\right}

\def\ket#1{\l|#1\rr\rangle}

\def\bnum{\begin{enumerate}}
\def\enum{\end{enumerate}}

\def\~{\hspace{0.5mm}}
\def\beq{\begin{equation}}
\def\eeq{\end{equation}}
\def\be{\begin{equation}}
\def\ee{\end{equation}}
\def\bea{\begin{eqnarray}}
\def\eea{\end{eqnarray}}
\def\half{\mbox{$1\over2$}}

\def\ba{{\bf a}}
\def\ba{{\bf d}}

\def\br{{\bf r}}

\def\cS{{\cal S}}
\def\ve{{\varepsilon}}
\def\cU{\mathcal{U}}
\def\cF{\mathcal{F}}
\def\tH{\tilde{H}}
\def\tU{\tilde{U}}
\def\tA{\tilde{A}}
\def\tB{\tilde{B}}
\def\cS{\mathcal{S}}
\def\cD{\mathcal{D}}
\def \proj{\mathcal{P}_\ell}
\def\avg#1{\l\langle#1\rr\rangle}

\def\hU{\hat{U}}

\makeatletter
 
 \@ifundefined{textcolor}{}
 {%
   \definecolor{BLACK}{gray}{0}
   \definecolor{WHITE}{gray}{1}
   \definecolor{RED}{rgb}{1,0,0}
   \definecolor{GREEN}{rgb}{0,1,0}
   \definecolor{BLUE}{rgb}{0,0,1}
   \definecolor{CYAN}{cmyk}{1,0,0,0}
   \definecolor{MAGENTA}{cmyk}{0,1,0,0}
   \definecolor{YELLOW}{cmyk}{0,0,1,0}
 }

\makeatother

\begin{document}
\title{The anomalous Floquet-Anderson insulator as a non-adiabatic quantized charge pump}
\author{Paraj Titum}
\affiliation{Institute for Quantum Information and Matter, Caltech, Pasadena, California 91125, USA}
\affiliation{Physics Department, Technion, 320003 Haifa, Israel}
\author{Erez Berg}
\affiliation{Department of Condensed Matter Physics, The Weizmann Institute of Science, Rehovot, 76100, Israel}
\author{Mark S. Rudner}
\affiliation{Niels Bohr International Academy and Center for Quantum Devices, University of Copenhagen, 2100 Copenhagen, Denmark}
\author{Gil Refael}
\affiliation{Institute for Quantum Information and Matter, Caltech, Pasadena,
California 91125, USA}
\author{Netanel H. Lindner}
\affiliation{Physics Department, Technion, 320003 Haifa, Israel}
\begin{abstract}
Periodically driven quantum systems provide a novel and versatile platform for realizing topological phenomena. 
 Among these are analogs of topological insulators and superconductors, attainable in static systems; however, some of these phenomena are unique to the periodically driven case. 
Here, we show that disordered, periodically driven systems admit an ``anomalous'' two dimensional phase,
whose quasi-energy spectrum consists of chiral edge modes that coexist with a fully localized  bulk - an impossibility for static Hamiltonians.
This unique situation serves as the basis for a new topologically-protected non-equilibrium transport phenomenon: quantized non-adiabatic charge pumping. 
We identify the bulk topological invariant that characterizes the new phase (which we call the ``anomalous Floquet Anderson Insulator'', or AFAI). 
We provide explicit models which constitute a proof of principle for the existence of the new phase. 
Finally, we present evidence that the disorder-driven transition from the AFAI to a trivial, fully localized phase is in the same universality class as the quantum Hall plateau transition.
\end{abstract}
\maketitle
\section{Introduction}
\label{sec: intro}
Time-dependent driving opens many new routes for realizing and studying topological phenomena in many-body quantum systems.
Recently, an intense wave of activity has developed around  
exploring the possibilities of using periodic driving to realize ``Floquet topological insulators'', i.e., driven system analogues of topological insulators \cite{Oka2009,Inoue2010,KBRD,Lindner2011,Lindner2013, Gu11,Kitagawa2011,Delplace2013,Podolsky2013,Titum2015,TorresPRB2014,TorresPRL2014,AlessioArxiv2014, Dehghani2014, Dehghani2014b, Bilitewski2014, Sentef2015,Seetharam2015, Iadecola2015} in a variety of solid state \cite{Wang2013}, atomic, and optical contexts~\cite{Jotzu2014,Rechtsman2013}.
Beyond these analogies, driven systems may also host their own unique types of robust topological phenomena, which have no analogues in non-driven systems~\cite{KBRD,Jiang2011,Rudner2013,Kundu2013,Carpentier,Asboth2014}. 
The latter will be at the heart of the present work.


In static (time-independent) two dimensional systems, the appearance of chiral edge states is intimately tied to the topological structure of the bulk Bloch bands, captured by the so-called Chern number~\cite{Thouless1982}. 
This well-known bulk-edge correspondence breaks down in periodically driven crystalline systems, where chiral edge states can exist even if the Chern numbers of all the bulk bands are zero~\cite{KBRD,Rudner2013}. Such anomalous edge states are captured instead by a topological invariant that characterizes the time evolution operator of the bulk wave functions~\cite{Rudner2013}.
A system exhibiting this anomalous behavior has been recently realized using microwave photonic networks \cite{Hu2015}.

\begin{figure}
\includegraphics[width=0.9\columnwidth]{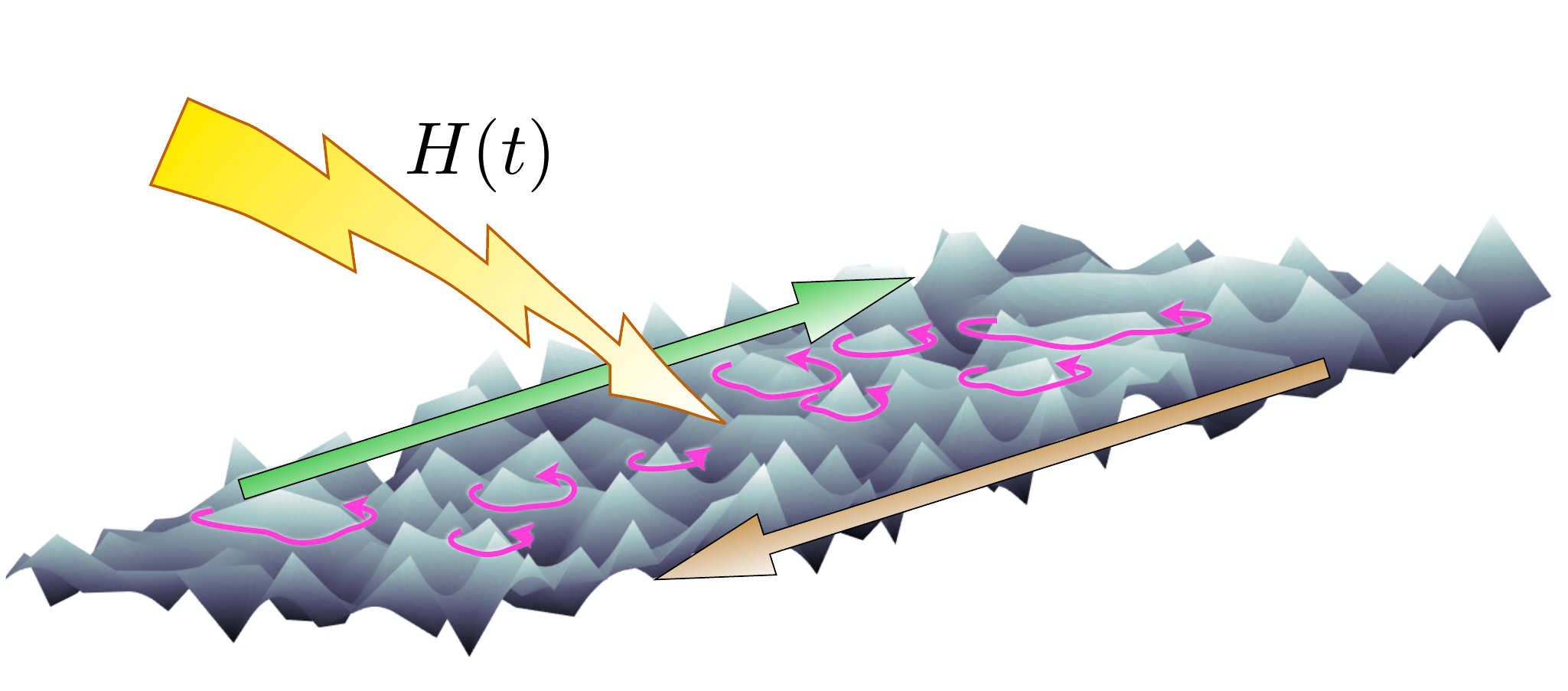}
\caption{The anomalous Floquet-Anderson insulator (AFAI), in a disordered two-dimensional periodically-driven system with time-dependent Hamiltonian $H(t)$.  In the AFAI phase all bulk states are localized, yet the system hosts chiral propagating edge states at all quasienergies.  The nontrivial topology of the phase is characterized by a nonzero value of the winding number defined in Eq.~\eqref{eq: invariant}}
\label{fig:intro}
\end{figure}

This unique topological phenomenon opens new possibilities, which are inaccessible in static systems. For example, it is well-known that Bloch bands with non-zero Chern numbers cannot be spanned by a complete basis of localized Wannier functions \cite{Thouless1984, Thonhauser2006}. Correspondingly, when quenched disorder is introduced, not all the states in a Chern band can be localized; delocalized states must exist at least at one value of energy in the band \cite{Halperin1982}. Intuitively, this can be understood from the fact that the chiral edge states in the bulk band cannot ``terminate'' without hybridizing with a delocalized bulk state. In contrast, a periodically driven system can exhibit chiral edge states even when all the Chern numbers are zero.  Moreover, due to the periodicity of quasienergy, it is in principle possible for a chiral edge state to wrap around the entire quasi-energy zone without terminating at a delocalized bulk state. This leads us to hypothesize that, in a disordered periodically-driven system, robust chiral edge states may coexist with an entirely localized bulk. In such a system, which we term an anomalous Floquet-Anderson insulator (AFAI), the chiral edge states  form a uni-directional one dimensional system, whose dynamics is decoupled from the bulk at all quasienergies. This situation defies the standard intution from strictly one-dimensional systems that must have an equal number of right and left moving modes. But can an AFAI state really exist? And if so, what are its physical consequences?

In this work, we explore the AFAI phase in periodically driven, two dimensional disordered systems. We construct explicit models that demonstrate its existence, and discuss its topological characterization and its physical properties. Strikingly, the AFAI hosts a unique \textit{non-equilbrium} topological transport phenomenon: quantized charge pumping in a non-adiabatic setting. Essentially, if all the states in the vicinity of the edge are occupied by fermions (to a distance of several times the bulk localization length), the uni-directional edge states carry a current whose long-time average is quantized in units of one particle per driving period. Importantly, disorder is essential for the quantization of pumping in the AFAI; absent the disorder, generically there is no quantization due to the presence of delocalized states in the bulk.

 Quantized pumping is well-known from the work of Thouless on adiabatically driven one-dimensional systems. however, unlike in the Thouless pump, in the AFAI, the driving frequency is not required to be small in order to observe the quantization of the current. The reason the current at the edge of an AFAI can remain quantized even when the adiabatic condition is violated is that the two counter-propagating edge modes that carry the current are \emph{spatially separated}~\cite{Thouless1983}; hence, they cannot backscatter into each other even if the driving frequency is not small.

 This paper is organized as follows.  In Sec.~\ref{sec: top invariant} we introduce the defining properties of the AFAI; we discuss the topological invariant characterizing the AFAI, and show that the AFAI exhibits edge modes at every quasi-energy. In Sec.~\ref{sec: charge pumping} we show how the edge mode structure leads to quantized charge pumping. We then demonstrate, in Sec.~\ref{sec: model}, the appearance and robustness of an AFAI in a simple, tractable model. In Sec.~\ref{sec: numerical results} we conduct a numerical study of a wider class of models exhibiting the AFAI phase. We numerically demonstrate the properties discussed in sections~\ref{sec: top invariant}~and~\ref{sec: charge pumping}. At strong disorder, we find a topological transition between the AFAI and a ``trivial'' Floquet insulator where {\it all} states are localized (including at the edges); we speculate that the transition is in the same universality class as the quantum Hall plateau transition, and corroborate  this using our numerical results.

\section{The AFAI: Topological Invariants and Edge States}
\label{sec: top invariant}
We begin by defining the AFAI, and introducing the topological invariant which characterizes it. The defining characteristic of the AFAI phase is the peculiar relationship between its bulk and edge mode spectra: in the AFAI phase all bulk Floquet states are localized, yet the system still hosts topologically-protected chiral modes along its edges.  

We consider a two-dimensional system of non-interacting particles with a time-periodic Hamiltonian, $H(t) = H(t+T)$, where $T$ is the driving period. No spatial translational symmetry is assumed. 
The interesting aspects of the AFAI phase are revealed by comparing the Floquet operators
$U(T)=\mathcal{T}e^{-i\int_0^T dt H(t)}$ for toroidal and cylindrical geometries. In an AFAI phase, all the eigenstates of $U(T)$ on a torus are localized. However, as we will show, in a cylindrical geometry there are eigenstates which are localized at the boundaries of the cylinder, but delocalized along the direction of the boundary, at \textit{every} quasi energy $0\leq \varepsilon  < \Omega\equiv 2\pi/T$.
The topological invariant which describes the AFAI is a generalization of the ``winding number'' introduced in Ref.~\cite{Rudner2013}. 
As a first step in constructing the topological invariant, 
we define an associated, time-periodic evolution operator for the system on a torus:
 \begin{equation}
 \cU_\ve(t)=U(t)\exp\left(iH_\ve^{\textrm{eff}}t\right),
 \label{eq: def U epsilon}
 \end{equation}
  with $H_\ve^{\textrm{eff}}=  \frac{i}{T}\log U(T)$. Note that, by construction, $\cU_\ve(T) = \mathds{1}$. The explicit dependence on $\ve$ in the above definitions comes from the 
necessary choice of a branch cut for $\log$; we use a definition such that $-i\log e^{i\chi}=\chi$ if $\chi\in \left[0,\ve T\right)$ and $-i\log e^{i\chi}=\chi-2\pi$ if $\chi \in \left[\ve T,2\pi\right)$. As an additional ingredient, we also consider a family of time-dependent Hamiltonians $H(\Theta,t)$ and the associated evolution operators $\cU( \Theta, t)$, in which constant (time independent) fluxes $\Theta=(\theta_x,\theta_y)$ are threaded through the torus \cite{footnote:gauge}.

With these definitions at hand, we can define the ``winding number''
\beq
W_\varepsilon=\int_0^T dt \int \frac{d^2\Theta}{8 \pi^2} \textrm{Tr}\left(\cU_\ve^\dagger\partial_t \cU_\ve \left[\cU_\ve^\dagger\partial_{\theta_x} \cU_\ve, \cU_\ve^\dagger\partial_{\theta_y} \cU_\ve\right]\right).
\label{eq: invariant}
\eeq
The winding number $W_\ve$ is an integer, which can in principle depend on the quasi-energy $\ve$. Note that in order for $W_\ve$ to be well defined, the quasi-energy $\ve$ has to remain in a spectral gap of $U(\Theta,T)$ for every value of the threaded fluxes $\Theta$ (otherwise, the operator $\cU_\ve$ is discontinuous as a function of $\Theta$). We argue that for a large enough system, almost all values of $\ve$ satisfy this requirement. This is because, upon changing fluxes $\theta_x$ and $\theta_y$, the quasi-energies of the localized bulk states only change by an amount proportional to $e^{-L/\xi}$, where $\xi$ is the localization length and $L$ is the linear system size. In contrast, the average level spacing is proportional to $1/L^2$. 

\begin{figure}[t]
\includegraphics[width=1.0 \columnwidth]{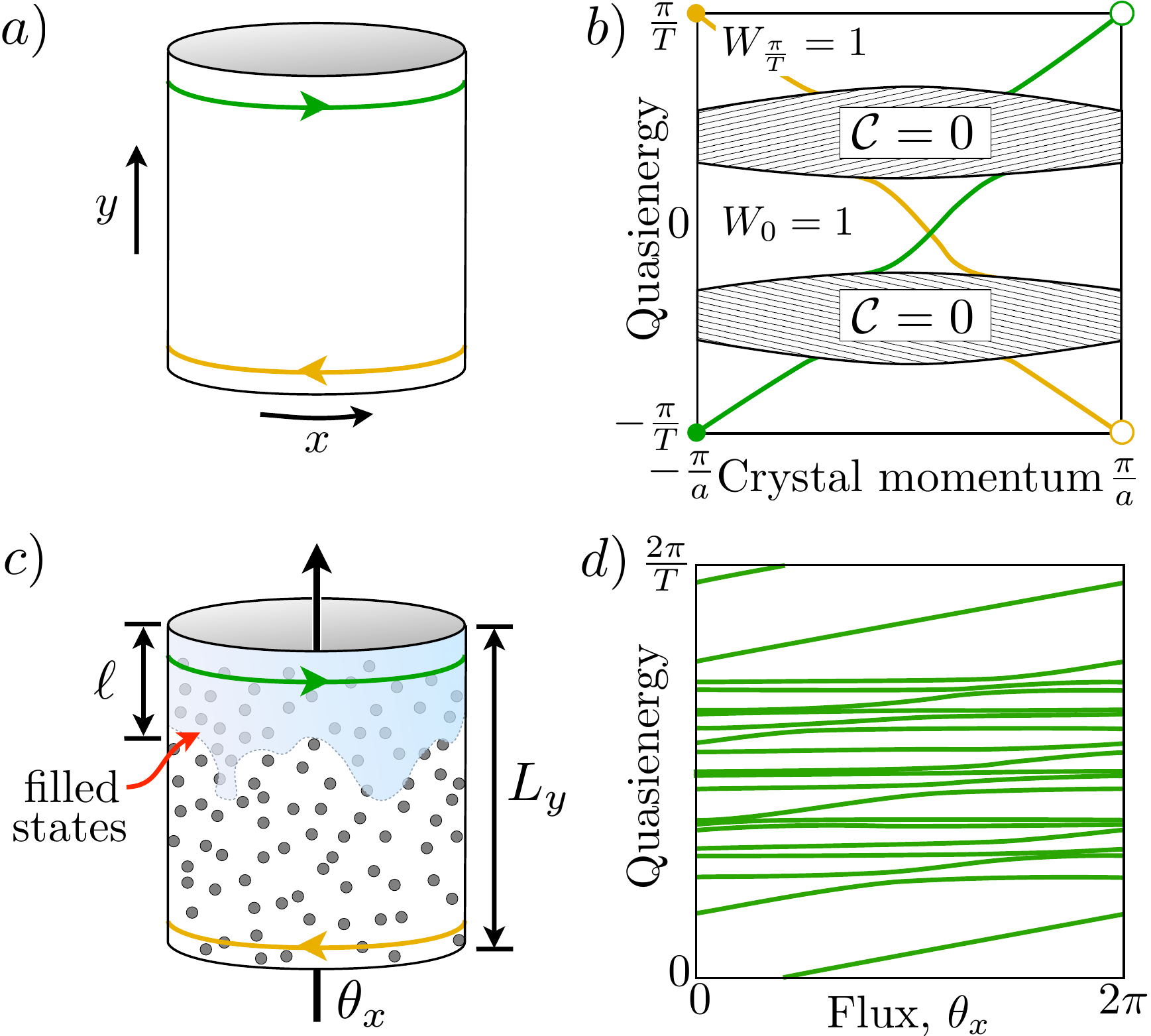}
\caption{Edge states and spectral flow in the AFAI. a) The parent phase of the AFAI is a clean system without disorder, where all Floquet bands have Chern number zero but the winding number (2) is non-zero in all gaps. In a cylinder geometry, chiral edge state propagate along the upper and lower boundaries, \textit{only} at quasi-energies  within the bulk gaps.   b) The corresponding spectrum, shown as a function of the conserved circumferential crystal momentum component.  c) When disorder is added, all bulk states become localized while the chiral edge modes on the cylinder persist.  When all states are filled near one end of the cylinder, a quantized current flows along the edge. d) With disorder, crystal momentum is no longer a good quantum number.  However, the spectrum of states localized near the upper edge, displayed as a function of the flux $\theta_x$ threaded through the cylinder, clearly displays a non-trivial spectral flow. The spectral flow \textit{fully} winds around the quasi-energy zone, accounting for the quantized pumping in the AFAI phase.
}
\label{fig: cylinder}
\end{figure}

Next, we show that if all the eigenstates of $U(T)$ are localized, then the invariant $W_\varepsilon$ is in fact independent of $\varepsilon$. 
This follows from the relation between the winding number $W_\varepsilon$ and the Chern numbers characterizing the eigenstates of $U(T)$~\cite{Rudner2013},
\begin{equation}
W_{\varepsilon_1}-W_{\varepsilon_2}= C_{\varepsilon_1,\varepsilon_2}.
\label{eq: chern diff}
\end{equation}
In the above, $C_{\varepsilon_1,\varepsilon_2}$ is the total Chern number of the eigenstates with quasi-energies between $\ve_1$ and $\ve_2$:
\begin{equation}
C_{\varepsilon_1,\varepsilon_2}=\int \frac{d^2\Theta}{4\pi} \textrm{Tr}\left\{P^{(\varepsilon_1,\varepsilon_2)}_\Theta\left[\partial_{\theta_x} P^{(\varepsilon_1,\varepsilon_2)}_{\Theta}, \partial_{\theta_y} P^{(\varepsilon_1,\varepsilon_2)}_{\Theta} \right]\right\},
\label{eq: yosi formula}
\end{equation}
where $P^{(\varepsilon_1,\varepsilon_2)}_\Theta$ is a projector onto the eigenstates of $U(\Theta,T)$ with quasi-energies between $\varepsilon_1$ and $\varepsilon_2$. If all the bulk eigenstates are localized, $C_{\ve_1, \ve_2} = 0$~\cite{note-Chern}. Therefore in this case by Eq.~(\ref{eq: chern diff}), $W_{\ve_1} = W_{\ve_2}$ for every pair of quasi-energies. Then, we can drop the subscript $\ve$, and refer to the winding number simply as $W$.

We thus define the AFAI as a time-periodic, disordered system in which (1) all the bulk Floquet eigenstates are localized, (2) the quasi-energy independent winding number $W$ is non-zero. Below, we argue that the boundaries of the AFAI necessarily support chiral edge states at every value of the quasi-energy.



Having defined the AFAI phase, we briefly mention how it can be reached. A good starting point for accessing the AFAI is a clean (translationally invariant) Floquet-Bloch system, for which all the Chern numbers of $U(T)$ vanish, but with $W_{\ve}  \neq 0$ for any $\ve$ within one of the gaps in the quasi-energy spectrum. A schematic quasi-energy spectrum of such a system in a cylindrical geometry is shown in Fig.~\ref{fig: cylinder}(b). We then add a static, spatially disordered potential to $H(t)$. 
We will argue below that all the bulk states are generically localized even for arbitrarily weak disorder. The crucial point is that \emph{the winding number $W$ need not be zero, even if all the bulk states are localized.} A specific solvable model, which serves as a proof of principle for the existence of the AFAI phase, is given in Sec.~\ref{sec: model}.


We finish this section with a discussion of the edge structure of the AFAI. In the clean limit, there are chiral edge states in any bulk quasi-energy gap with a non-zero winding number~\cite{Rudner2013}. Clearly, these edge states cannot localize when disorder is added. Moreover, intuitively, if all the bulk states are localized, the chiral edge states must persist even within the bulk bands. To see this, consider a system in a cylindrical geometry. Upon inserting a flux quantum through the hole of the cylinder, the chiral edge states exhibit a non-trivial ``spectral flow'': i.e., even though the spectrum as a whole is periodic as a function of flux, every state evolves into the next state in the spectrum [Fig.~\ref{fig: cylinder}(d)]. The spectral flow cannot terminate in the bulk bands. Since all the bulk states are localized, they are insensitive to the flux, and hence there must exist  a delocalized, chiral edge state at every quasi-energy within the bulk bands that ``carry'' the spectral flow.


To make this argument more precise, we define a topological invariant that directly characterizes the spectral flow of the edge states. This topological invariant turns out to be equal to the bulk invariant $W$; we will show this in detail in Sec.~\ref{sec: charge pumping} and Appendix~\ref{appendix: current}, where we demonstrate that both invariants are related to quantized charge pumping along the edge.

To construct the edge topological invariant, we consider a cylinder that extends from $y=0$ to $y=L_y$, with a flux $\theta_x$ inserted through the hole of the cylinder. 
The evolution operator of the system on the cylinder is denoted by $\tilde{U}(t)$. We now isolate the topological features of the edge states by deforming the evolution operator in the regions away from the edges, such that the evolution in the bulk takes a simple universal form, while the evolution near the edges is unaffected.  In particular, we “flatten” the bulk evolution such that all Floquet eigenstates localized sufficiently far (at least a distance $\ell_0$) from the edges have quasienergy $\varepsilon = 0$. The resulting evolution operator interpolates smoothly between $\tilde{U}(t)$ in the vicinity of the edge and $\mathcal{\tilde{U}}_\ve(t)$ of Eq.~(\ref{eq: def U epsilon}) in the bulk (an explicit formulation of the deformation procedure appears in Appendix~\ref{app:flattening}).  The deformed evolution operator takes a block-diagonal form:
\begin{equation}
\mathcal{\tilde{U}}_{\ve}(T)=
\begin{pmatrix}
\mathcal{\tilde{U}}_1(T) &0&0\\
0 & \mathds{1}  &0 \\
0 & 0 & \mathcal{\tilde{U}}_2(T)
\end{pmatrix},
\label{eq: U matrix}
\end{equation}
where in the above, the sub-blocks  $\mathcal{\tilde{U}}_1(T)$ and $\mathcal{\tilde{U}}_2(T)$ correspond to sites with $0\leq y\leq \ell_0$ and $L_y-\ell_0 \leq y \leq L_y$, respectively; the unity block acts on sites with $\ell_0 < y < L_y-\ell_0$. The precise value of $\ell_0$ is not important, as long as it is much larger than the bulk localization length of the original evolution operator, $\tilde{U}(T)$.
The integer-valued ``edge winding number'' is defined as
\begin{eqnarray}
n_{\mathrm{edge}} &=& \int_0^{2\pi} \frac{d\theta_x}{2\pi} \mathrm{Tr}\left[ \mathcal{\tilde{U}}_1(T)^\dagger \partial_{\theta_x}\mathcal{\tilde{U}}_1(T)\right] \nonumber \\
&=&\sum_{j} \frac{T}{2\pi} \int_0^{2\pi} d\theta_x \frac{\partial \ve_j}{\partial \theta_x},
\label{eq:nedge}
\end{eqnarray}
where the sum in the second line runs over all the eigenstates of $\mathcal{\tilde{U}}_1(T)$, and $\ve_j$ are their corresponding eigenvalues. The edge winding number (\ref{eq:nedge}) 
counts how many times the spectrum of $\mathcal{\tilde{U}}_1(T)$ ``wraps'' around the quasi-energy zone, $\ve \in [0,2\pi/T)$, as $\theta_x$ varies from $0$ to $2\pi$. A schematic example of a spectrum with a non-zero winding number is shown in Fig.~\ref{fig: cylinder}(d). Note that the \emph{total} winding number of the system, $\int_0^{2\pi} \frac{d\theta_x}{2\pi} \mathrm{Tr}\left[ \mathcal{\tilde{U}}_\ve(T)^\dagger \partial_{\theta_x}\mathcal{\tilde{U}}_\ve(T)\right]$, must vanish~\cite{KBRD}. Hence, the winding numbers of $\mathcal{\tilde{U}}_1(T)$ and $\mathcal{\tilde{U}}_2(T)$ must sum to zero.

A non-zero $n_{\mathrm{edge}}$ necessarily implies that there are delocalized states along the edge; if all states were localized, their quasi-energies would be almost insensitive to $\theta_x$, and hence $n_{\mathrm{edge}}$ would be zero. Note also that, since in the AFAI all the bulk states are localized, changing $\ell_0$ would not change $n_{\mathrm{edge}}$; this amounts to adding a few \emph{localized} states to the spectrum of $\mathcal{\tilde{U}}_1(T)$, and cannot change its winding number.

\section{Quantized charge pumping}
\label{sec: charge pumping}

We now discuss the physical implications of the AFAI phase. Consider an AFAI placed in a cylindrical geometry, as in Fig~\ref{fig: cylinder}(c). Fermions are loaded into the system such that in the initial state all the lattice sites are filled up to a distance of $\ell \gg \xi$ from one edge of the cylinder, and all the other sites are empty. Below we show that in the thermodynamic limit, the current across a vertical cut through the cylinder, averaged over many driving periods, is equal to
$n_{\mathrm{edge}}$, Eq.~(\ref{eq:nedge}), 
divided by the driving period $T$. The exact form in which we terminate the filled region will not matter, as long as all the sites near one edge are filled, and all the sites near the other edge are empty. The system thus serves as a quantized charge pump, but unlike the quantized pump introduced by Thouless~\cite{Thouless1983}, there is \emph{no requirement for adiabaticity}.

In Appendix~\ref{appendix: winding} we furthermore show by a direct evaluation that the long-time average of the pumped charge per driving period is also equal to $W$, the bulk invariant. In particular, this implies that $W = n_{\mathrm{edge}}$.

To set up the calculation of the charge pumping in the AFAI, we choose coordinates such that $x$ is the direction along the edges of the cylinder, and $y$ is the transverse direction. 
We denote the initial many-body (Slater determinant) state, in which all sites up to a distance of $\ell$ from the edge are filled, 
by $|\Psi (0)\rangle$. Then, the charge pumped across the line $x=x_0$ between $t=0$ and $t=\tau$ is given by
\begin{equation}
\langle Q \rangle_{\tau}= \int_0^\tau dt \Big\langle\Psi(t)\Big| \frac{\partial \tH(\theta_x,t) }{\partial \theta_x}\Big|\Psi(t)\Big\rangle.
\label{eq: current def}
\end{equation}
Here, $\theta_x$ is the flux through the cylinder and $\tH(\theta_x,t)$ is the corresponding Hamiltonian. For Eq.~\eqref{eq: current def}, we use a gauge such that on the lattice, every hopping matrix element that crosses the line $x=x_0$ has a phase of $e^{i\theta_x}$.

The initial state $|\Psi (0)\rangle$ clearly does not return to itself after a single driving period. Therefore, we cannot expect that the pumped current to be identical between different periods along the evolution, nor can we expect it to be exactly quantized. However, we find that the average pumped charge over $N$ periods approaches a quantized value $Q_\infty$ in the limit of a large number of periods, where the correction to the quantized value decays as $1/N$:
\begin{equation}
\frac{\langle Q \rangle_{NT}}{N}=Q_\infty+O\left(\frac{1}{N}\right).
\label{eq: Qinfty def}
\end{equation}
Here, $Q_\infty = W = n_{\mathrm{edge}}$ (where $W$, $n_{\mathrm{edge}}$ are the bulk and edge topological invariants, respectively, defined in Sec.~\ref{sec: top invariant}). Note that  $Q_\infty$ is independent of $x_0$, i.e.,  the  charge pumped across any line parallel to the $y$ axis leads to the same $Q_\infty$.

In order to compute the charge pumped per period, it is useful to express $|\Psi (t)\rangle$ as a superposition of the Floquet eigenstates. 
As we show in Appendix~\ref{appendix: spectral flow}, when averaging the pumped charge over $N$ periods, the contribution of the off-diagonal terms between different Floquet eigenstates decays at least as fast as $1/N$. The diagonal terms yield a contribution that depends on the evolution over a \textit{single} period, giving
\begin{equation}
Q_\infty = \sum_j n_j \int_0^T dt \langle \psi_j(t)| \frac{\partial \tH(\theta_x,t) }{\partial \theta_x}|\psi_j(t)\rangle.
\label{eq: Q infty}
\end{equation}
In the above, $|\psi_j(t)\rangle$ are the single particle Floquet states, which evolve in time as $|\psi_j(t)\rangle=e^{-i\ve t}|\phi_j(t)\rangle$ (where $|\phi_j(t)\rangle$ is periodic in time), and $n_j$ are the Floquet state occupation numbers in the initial state, $n_j=\langle \Psi(0) |\psi^\dagger_j \psi^{\vphantom{\dagger}}_j |\Psi (0)\rangle$, where $\psi_j^\dagger$ is the creation operator corresponding to $|\psi_j(0)\rangle$. [Note that if fermions were initialized  in the Floquet eigenstates, such that $n_j=0$ or $n_j=1$, we would obtain $\avg{Q}_{NT}/N=Q_\infty$, without the correction terms in Eq.~\eqref{eq: Qinfty def}].

Straightforward manipulations yield $Q_\infty= T\sum_j n_j \partial\ve_j/\partial\theta_x$. At this point, the average current per period depends on $\theta_x$. In the thermodynamic limit, we expect this dependence to disappear. As in the case of the quantization of the Hall conductance \cite{Avron1985}, 
we average over $\theta_x$~\cite{Spiros2015-comment}. We therefore get
\begin{equation}
Q_\infty= \frac{T}{2\pi}\sum_j \int_0^{2\pi}d\theta_x  n_j  \frac{\partial\ve_j}{\partial\theta_x}.
\label{eq: current spec flow}
\end{equation}

Equation~(\ref{eq: current spec flow}) relates the average current in a period to the spectral flow of the Floquet spectrum as the flux $\theta_x$ is threaded. It is reminiscent of the expression for the edge topological invariant, $n_{\mathrm{edge}}$, Eq.~(\ref{eq:nedge}), defined in terms of the ``deformed'' evolution operator $\mathcal{\tilde{U}}_\ve (T)$. Below, we give a heuristic argument that indeed $Q_\infty = n_{\mathrm{edge}}$, up to corrections that are exponentially small in $\ell$. A more rigorous (but technically cumbersome) derivation of the relation between the pumped charge and the bulk invariant is presented in Appendix~\ref{appendix: current}. Numerical evidence for the quantization of the pumped charge is shown in Sec.~\ref{sec: numerical results}. 

Our strategy in analyzing the pumped charge is to deform the evolution operator into the ``ideal'' form, $\mathcal{\tilde{U}}_\ve (T)$ of Eq.~(\ref{eq: U matrix}), for which the pumped charge is exactly quantized, and to put bounds on the correction to the pumped charge due to the deformation. We define the deformation process according to Appendix~\ref{app:flattening}, with $\ell_0$, the width of the strip beyond which the quasi-energy spectrum becomes flat, chosen such that $\ell \sim \ell_0$. 
Clearly, for the deformed evolution operator, $n_j = 1$ for every eigenstate of $\mathcal{\tilde{U}}_1$. Therefore, the deformed evolution operator has an exactly quantized pumped charge, equal to $n_{\mathrm{edge}}$.

Now, consider the pumped charge of the original (undeformed) evolution. We can roughly divide the Floquet states that contribute to Eq.~(\ref{eq: current spec flow}) into three categories:

\begin{enumerate}

\item{States that are localized far from occupied region, $y \gg \ell$. For these states, $n_j$ is exponentially small, and hence their contribution to $Q_\infty$ is negligible.}

\item{States that are localized near the edge, $y \ll \ell$. These states have $n_j \approx 1$. Their wavefunctions and quasi-energies, and hence their contribution to $Q_\infty$, are essentially unaffected by the deformation process.}

\item{States that are localized near the boundary between occupied and unoccupied sites, $y \sim \ell$. For such states, $n_j$ is neither close to $0$ nor to $1$; however, 
these states are localized in the $x$ direction (as are all the bulk states in the AFAI). Therefore, $\partial \ve_j / \partial \theta_x$ of these states is exponentially small, and they contribute negligibly to $Q_\infty$.}

\end{enumerate}

As $\theta_x$ varies, there are avoided crossings in the spectrum, in which the character of the eigenstates changes. E.g., an eigenstate localized around $y_1 \ll \ell$ may undergo an avoided crossing with an eigenstate localized around $y_2 \sim \ell$. When $\theta_x$ is tuned to such degeneracy points, the two eigenstates hybridize strongly, and do not fall into either of the categories discussed above. Such resonances affect both ${\partial\ve_j}/{\partial\theta_x}$ and the occupations $n_j$ of the resonant states. However, since the eigenstates that cross are localized in distant spatial areas, the matrix element that couples them is exponentially small. Therefore significant hybridization requires their energies to be tuned into resonance with exponential accuracy, limiting the regions of deviation to exponentially small ranges of $\theta_x$, of order $e^{-\ell/\xi}$. The number of such resonances increases only polynomially with the size of the system, and therefore for $L_y \gg \ell \gg  \xi$ and $L_x \propto L_y$, their effect on $Q_\infty$ is exponentially small.

We conclude that, in the thermodynamic limit, all the contributions to $Q_\infty$ in Eq.~(\ref{eq: current spec flow}) that are not exponentially suppressed are also exponentially insensitive to the deformation process. Therefore, $Q_\infty = n_{\mathrm{edge}}$.

\section{Model for an anomalous Floquet-Anderson phase}
\label{sec: model}
In this section, we study a 
simple model which allows us to explicitly demonstrate 
the existence and robustness of the AFAI phase.
We start from a solvable model introduced in Ref.~\cite{Rudner2013}, which exhibits perfectly flat bulk Floquet bands, and hosts chiral edge modes at its boundaries. 
Adding a specific kind of disorder to this model results in localization of all the bulk states, while preserving the edge states; the system is thus in the AFAI phase. We then argue that this phase is robust to generic small perturbations (i.e., the bulk states remain localized, and the chiral edge states persist).

\begin{figure}[t]
\includegraphics[width=0.9 \columnwidth]{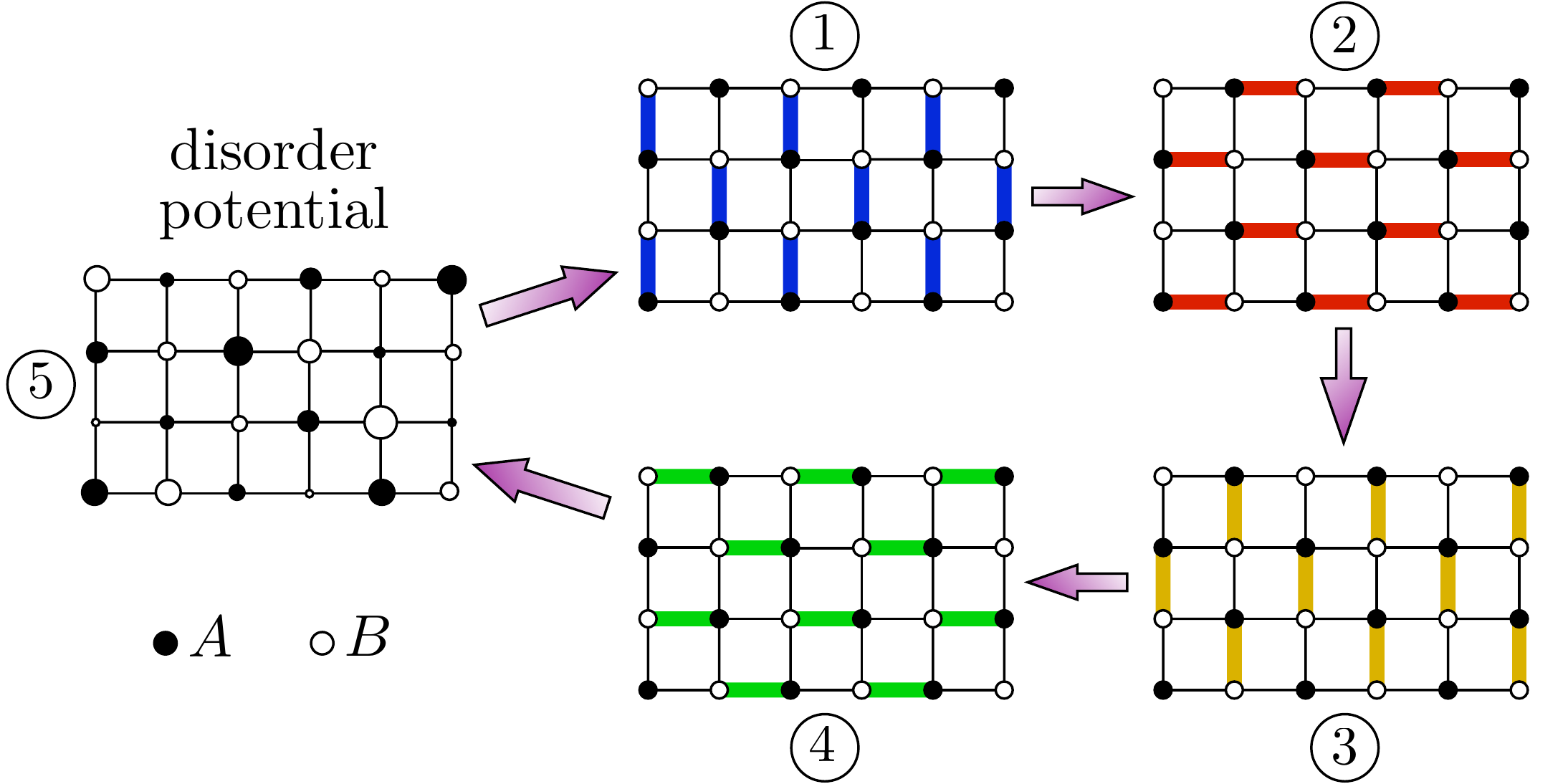}
\caption{Simple model for achieving the anomalous Floquet-Anderson phase.
The Hamiltonian is piecewise constant, defined in five equal length segments of duration $T/5$.
During steps 1-4, nearest-neighbor hopping is applied along the colored bonds as shown.
The hopping strength $J$ is chosen such that a particle hops between adjacent sites with probability one during each step.
In the fifth step, all hopping is turned off and a random disorder potential is applied (the same potential is used for all subsequent driving cycles).}\label{fig:model}
\end{figure}

We consider a system on a square lattice with a periodic, piecewise-constant Hamiltonian of the form: 
$H_{\mathrm{clean}}(t)=H_{n}$, for $\frac{(n-1)T}{5}\le t<\frac{nT}{5}$, $n=1,\dots,5$.
The square lattice is divided into two sublattices, $A$ and $B$ (shown as filled and empty circles in Fig.~\ref{fig:model}, respectively).
During each of the first four segments of the driving, $n = 1,\dots,4$, 
hopping matrix elements of strength $J$ between the $A$ and $B$ sublattices are turned on and off in a cyclic, clockwise fashion, as shown in Fig.~\ref{fig:model}: during segment $n=1$, $2$, $3$, or $4$, each site in the A sublattice is connected by hopping to the site above, to the right, below, or to the left of it, respectively.
In the fifth segment of the period, all the hoppings are set to zero, and an on-site potential $\delta_{A,B}$ is applied on the $A$ and $B$ sublattice sites, respectively.

We choose the hopping strength $J$ such that $\frac{JT}{5}=\frac{\pi}{2}$.
For this value of $J$, during each hopping segment of the driving period
a particle that starts on one of the sites hops to the neighboring site with unit probability. The on-site potential, applied only while all hopping matrix elements are turned off, is chosen to be
$\delta_{A,B}=\pm\frac{\pi}{2T}$. With this time-dependent Hamiltonian,
it is easy to find the Floquet eigenstates and quasi-energies.
The bulk spectrum consists of 
two flat Floquet bands with quasi-energies $\pm\frac{\pi}{2T}$, 
with the corresponding eigenstates localized on either the $A$ or $B$ sublattice.
The winding number invariant can be computed for this model at $\ve=0,\pi/T$, yielding $W_0=W_{\pi/T}=1$~\cite{Rudner2013}.
In a cylindrical geometry the two edges host linearly dispersing chiral modes in the quasi-energy gaps between the two bulk bands. 

We now introduce a specific form of a 
time-dependent disorder potential, $V(t)$, which still allows for an exact solution.
The full time-dependent Hamiltonian is given by $H_0(t) = H_{\mathrm{clean}}(t) + V(t)$.
During the fifth segment of the driving period, we let $V_j(t) = \sum_j V_{j}c^\dagger_jc_j$, where
$V_{j}$ is a uniformly distributed in the range $[-V,V]$, and $c_j$  is the annihilation operator on site $j$.
During segments 1--4, $V_(t) = 0$.
We choose $V<\frac{\pi}{2T}$.

By following the evolution of a state that is localized on a single bulk site $j$ at time $t=0$, one can easily verify that this state is a Floquet eigenstate, whose quasi-energy is $\pm \frac{\pi}{2T} + V_j$ [here $+1$ ($-1$) refers to a site in the $A$ ($B$) sublattice].
The Floquet spectrum consists of two bands, with quasi-energies in the range $[\pm\frac{\pi}{2 T}-V,\pm\frac{\pi}{2 T}+V]$.
One can similarly follow the evolution of a state that is initially localized on a site at the edge; in the geometry of Fig.~\ref{fig:model} (viewed as a ``strip'' geometry with edges parallel to the horizontal axis), a state initialized on the $A$ ($B$) sublattice at the top (bottom) edge moves by two lattice constants to the right (left) every driving period.
Therefore, chiral edge states persist 
even in the presence of the disorder potential.

As long as the gaps in the quasi-energy spectrum at $\ve=0, \pi/T$ remain open, the winding numbers at these gaps cannot change.
Therefore, at least over a finite range of the disordered potential strength, we have $W_0=W_{\pi/T}=1$ as in the clean limit~\cite{Rudner2013}. Moreover, in the disordered system, all the bulk Floquet states are localized. Therefore, as argued in the previous section, the winding number is actually independent of the quasi-energy: $W_\ve = 1$ for all $\ve$. We conclude that by the definition presented in Sec.~\ref{sec: top invariant}, the Hamiltonian $H_0(t) = H_{\mathrm{clean}}(t) + V(t)$ realizes the AFAI phase.



Clearly, the above model utilizes a very specific form of the periodic driving and of the added disorder. 
Nevertheless, we argue that the AFAI is a robust phase that does not require fine-tuning.
To demonstrate the robustness of the phase, we now consider a generic local perturbation of $H_0(t)$ that
preserves the periodicity in time, $H_\lambda(t)=H_0(t)+\lambda D(t)$, and show that the AFAI phase survives up to a finite value of $\lambda$. 

The perturbation $D(t)$ is assumed to be periodic in time and short-ranged in real space, such that the matrix elements of $D(t)$ vanish beyond the $r$th neighbor on the square lattice. For $V=0$ (no disorder) and $\lambda \ne 0$, the bulk eigenstates of $U(T)$ are generically dispersive and delocalized. 
However, we argue that for
$V>0$ and for a sufficiently small $\lambda$, all the bulk Floquet states
remain localized.
To see this, we derive a time-independent effective Hamiltonian $H^{\rm eff}_\lambda$ for the 
Floquet problem (on the torus) with $V\ne0,\lambda\ne0$:
\begin{equation}
e^{-i H_\lambda^\mathrm{eff}T}=\mathcal{T} e^{-i\int_0^T\! dt\,\left[ H_0(t) + \lambda D(t)\right]},
\label{eq: static heff}
\end{equation}
where $\mathcal{T}$ denotes time ordering. We further write the effective Hamiltonian as $H_\lambda^{\mathrm{eff}} = H^{\mathrm{eff}}_{(0)}+ D_{\mathrm{eff}}$, where $H^{\mathrm{eff}}_{(0)}$ corresponds to the unperturbed ($\lambda = 0$) effective Hamiltonian, defined such that its eigenstates lie in the range $[-\frac{\pi}{T}, \frac{\pi}{T})$. 
Here, we are considering a system with periodic boundary conditions; we will comment on the edge states later.

The key point, which we show below, is that 
for sufficiently small $V$ and $\lambda$,  the hopping matrix elements of the effective
static Hamiltonian $H^{\mathrm{eff}}_{(0)} + D_{\mathrm{eff}}$ decay exponentially with distance. If, in addition, $\lambda \ll V/\Omega$, then all of the Floquet eigenstates
remain localized~\cite{Anderson1958}. 

To find the effective Hamiltonian for $\lambda\neq0$ we need to solve for $D_{\mathrm{eff}}$. The unperturbed effective Hamiltonian is of the form 
\begin{equation}
H^{\mathrm{eff}}_{(0)}=\sum_{j}\left(\frac{(-1)^{\eta_{j}}\pi}{2T}+V_{j}\right)c_{j}^{\dagger}c_{j}^{\vphantom{\dagger}}\label{eq:Heff0},
\end{equation}
where $\eta_{j}=0(1)$ for $j$ on the $A(B)$ sublattice.
We express $D_{\mathrm{eff}}$ as a power series in $\lambda$,
\begin{equation}
D_{\mathrm{eff}}=\lambda D_{\mathrm{eff}}^{(1)}+\lambda^2 D_{\mathrm{eff}}^{(2)}+\lambda^{3}D_{\mathrm{eff}}^{(3)}+\dots.
\label{eq:Deff}
\end{equation}
To find $D_{\mathrm{eff}}^{(m)}$ 
we expand both sides of Eq.~(\ref{eq: static heff}) in powers
of $\lambda$, and compare them order by order. The details of the calculation are given in Appendix~\ref{appendix: perturbation}. The results can be summarized by considering the explicit representation of the  operators $D_{\mathrm{eff}}^{(n)}$ as a tight-binding ``Hamiltonian,'' 
\begin{equation}
D_{\mathrm{eff}}^{(n)}=\sum_{i,j}\Delta_{ij}^{(n)}c_{i}^{\dagger}c_{j}.
\label{eq:tightbind}
\end{equation}
Using the explicit form for  $H^{\mathrm{eff}}_{(0)}$ given in Eq.~(\ref{eq:Heff0}), we find for the lowest-order term 
\begin{align}
\Delta_{ij}^{(1)} & =\frac{iE_{ij}}{e^{iE_{ij}T}-1}\left[\int_{0}^{T}\!dt\, \cD(t) \right]_{ij},
\label{eq: delta0}
\end{align}
with $\cD(t) = U_0(t,0)^\dagger D(t) U_0(t,0)$, where $U_0(t,0)$ is the $\lambda=0$ evolution operator, and $E_{ij}=V_{i}-V_{j}+\frac{\left[(-1)^{\eta_{i}}-(-1)^{\eta_{j}}\right]\pi}{2T}$
is the zeroth-order quasi-energy difference between the states localized at {sites} $i$ and $j$.

As long as $E_{ij} T$ is smaller than $2\pi$
for every pair of sites (which is the case for $V<\frac{\pi}{2T}$), the factor $\frac{iE_{ij}}{e^{iE_{ij}T}-1}$ in Eq.~(\ref{eq: delta0}) is bounded. Similarly, the matrix elements $\Delta^{(n)}_{i,j}$ are all non-singular (see Appendix~\ref{appendix: perturbation}). Under these
conditions, we expect the expansion in powers of $\lambda$ to converge. In Appendix~\ref{appendix: perturbation}, we argue that for sufficiently small $\lambda$, the matrix elements of $H^{\rm eff}_\lambda$ decay exponentially with distance. Therefore, $H^{\rm eff}_\lambda$ has the form of a tight-binding model with random on-site potentials and weak, short-range hopping.
{In this context,} we expect all states to remain localized up to a critical strength of $\lambda$.


Since all the bulk states remain localized as $\lambda$ is turned on, the chiral edge states that exist for $\lambda=0$ cannot disappear; the only way to remove them is by closing the mobility gap in the bulk, allowing the two counter-propagating states at the two opposite edges to backscatter into each other. Hence, we expect the edge chiral states, {and the associated quantized pumping},
to persist up to a critical value of $\lambda$ where the bulk mobility gap closes.

%

\section{Numerical results}
\label{sec: numerical results}

Numerical simulations substantiate the conclusions of Sections \ref{sec: top invariant}--\ref{sec: model}. We will first briefly summarize our main findings, and then describe the simulations and results in more detail in the subsections below. For the simulations, a variant of the model discussed in Sec.~\ref{sec: model}, defined on a square lattice, is used:
\begin{equation}
\tilde{H}(t) = H_{\mathrm{clean}}(t) + \lambda D + \sum_j V_j c_j^\dagger c_j^{\vphantom{\dagger}},
\label{eq: H numerics}
\end{equation}
where $H_{\mathrm{clean}}(t)$ is the time-dependent, piecewise-constant Hamiltonian described in Sec.~\ref{sec: model} (pictured in Fig.~\ref{fig:model}).  Using numerics, we are now able to study the more generic case in which the sublattice potential (denote here by $D$), as well as the disorder potential are \emph{time-independent} (in contrast to the model studied in Sec.~\ref{sec: model}).  We define $D=\frac{1}{2T}\sum_{j}(-1)^{\eta_{j}}c_{j}^{\dagger}c^{\vphantom{\dagger}}_{j}$, and take $V_j$ to be uniformly distributed in the interval $\left[-V,V\right]$. The parameters of the model are chosen to be $\lambda=\pi$, and $\delta_{AB}=0$.

In the clean case ($V=0$), the system exhibits an anomalous Floquet-Bloch band-structure: the Chern numbers of all the bulk bands are zero, but the winding number $W_\ve = 1$ for any value of $\ve$ within each of the band gaps~\cite{Rudner2013}. Such a band-structure is depicted in Fig.~\ref{fig: cylinder}(b).
When the disorder potential is turned on, however, the system enters the AFAI phase. Below, we show numerically that the bulk states become localized, and coexist with edge states which occur in all quasi-energies. Furthermore, when the system is initialized with fermions filling all of the sites in the vicinity of one edge, while the rest remain empty, as in Sec.~\ref{sec: charge pumping}, the disordered system exhibits quantized amount of charge pumped per period, when averaged over long times. Finally, we examine the behavior of the system as the strength of the disorder potential is increased.  We find that when the disorder strength reaches a certain critical value, the system undergoes a topological phase transition where the winding number changes from $1$ to $0$. For stronger disorder, a ``trivial'' phase (where all bulk states are localized and there are no chiral edge states) is stabilized.


\subsection{Localization, edge modes, and quantized charge pumping in the AFAI}
\label{sec: numerics weak}


The localization properties of the bulk Floquet eigenstates of (\ref{eq: H numerics}) can be extracted from the statistics
of the spacings between the quasi-energy levels. 
For localized states, the distribution of the level-spacing is expected to have a Poissonian form. In contrast, extended states exhibit level
repulsion and obey Wigner-Dyson statistics \cite{Mehta2004}.
To distinguish between these distributions, 
it is convenient to use the ratio between the spacings of adjacent quasi-energies levels~\cite{OganesyanHuse, Atas2013, DAlessio2014}. Choosing the quasi-energy zone to be between $-\pi/T$ and $\pi/T$ (i.e., choosing $-i\log e^{i\ve T}=\ve T$ for $-\pi/T\leq\ve<\pi/T$), we label quasi-energies in ascending order. We then define the level-spacing ratio (LSR) as $r=\min\left\{ \delta_{n},\delta_{n+1}\right\} /\max\left\{ \delta_{n},\delta_{n+1}\right\} $,
where $\delta_{n}=\ve_{n}-\ve_{n-1}$. This ratio, $r\le1$,
converges to different values for extended and localized states,
depending on the symmetries of the system. For localized
states, $r_{{\rm loc}}\approx 0.39$ \cite{OganesyanHuse}, while for
extended states, $r_{{\rm ext}}\approx0.6$~\cite{DAlessio2014}. 
The latter value is obtained when one assumes that the quasi-energies are distributed according to the circular unitary ensemble (CUE)~\cite{DAlessio2014}, and in the thermodynamic limit, coincides with the value obtained by the more familiar Gaussian unitary ensemble (GUE).

\begin{figure}
\includegraphics[width=1.0\columnwidth]{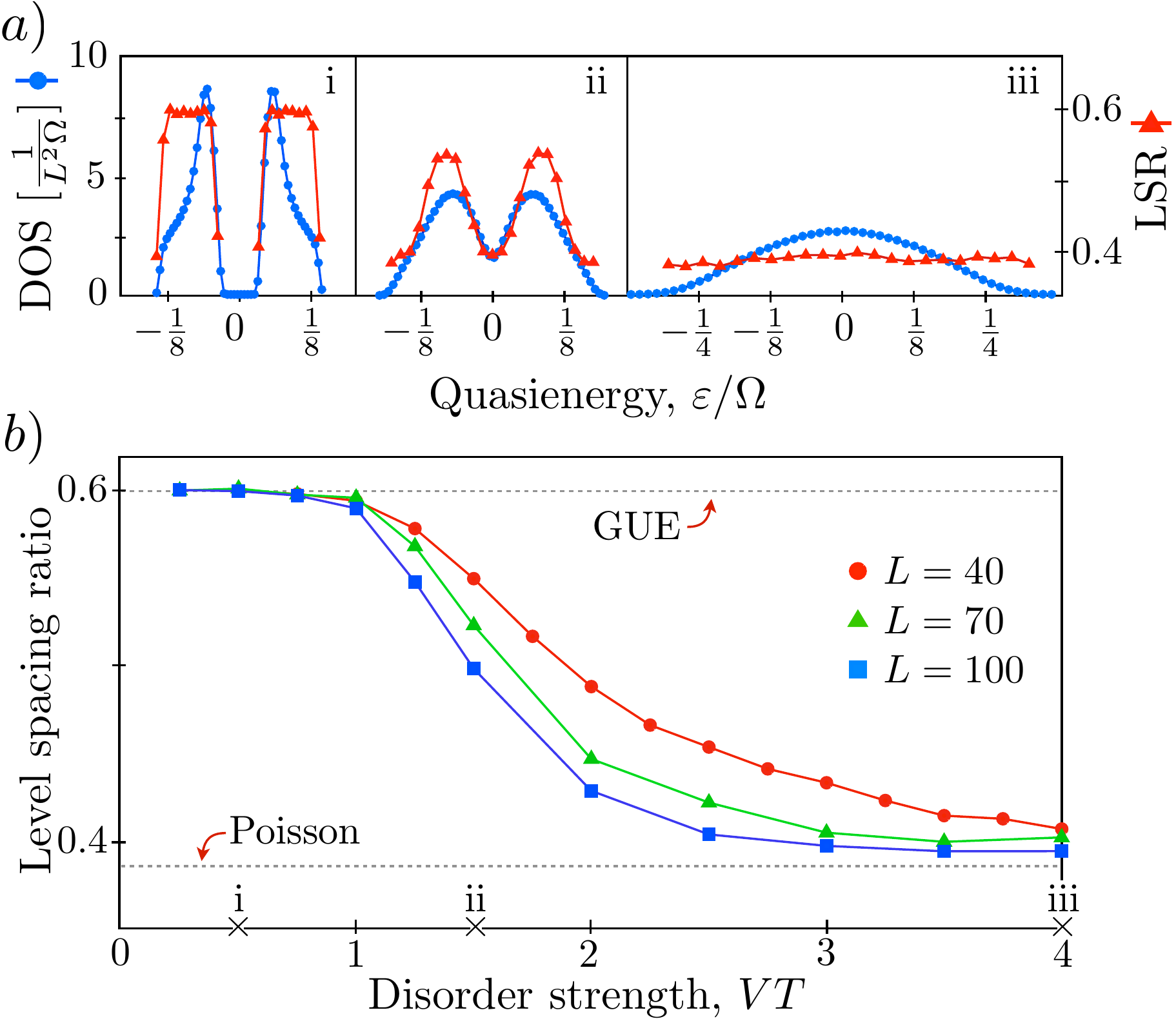}
\caption{
 Localization of Floquet states in the AFAI as a function of disorder strength, computed for the model presented in Eq.~\eqref{eq: H numerics}. We use $\lambda=\pi$ and an $L \times L$ system with periodic boundary conditions.
(a) Quasienergy density of Floquet states per unit area (DOS) and level spacing ratio (LSR), for three values of disorder strength as indicated by the markers on the axis of panel (b).
For all cases we take $L = 70$.
(b) Finite size scaling of the localization transition.
Level statistics in the delocalized regime are described by the Gaussian unitary ensemble (GUE), characterized by an average level spacing ratio $r_{\rm ext} \approx 0.60$; in the localized regime, Poissonian level statistics give $r_{\rm loc} \approx 0.39$.
These characteristic values are indicated by dashed lines. }
\label{fig:smalldisordertrans}
\end{figure}
Since the Floquet problem does not possess any generic symmetries such as time-reversal, particle-hole, or chiral symmetry, we expect its localization properties to be similar to those of the unitary class~\cite{Mirlin2008,AltlandZirnbauer1997,Dyson1962}. 
In analogy with the situation in static Hamiltonians in the unitary class~\cite{Pruisken1985}, we expect that arbitrarily weak disorder is sufficient to localize the all Floquet states (on the torus). However, for weak disorder, the characteristic localization length $\xi$ 
can be extremely long, and easily exceeds the system sizes accessible in our numerical simulations. Therefore, the level spacing ratio is expected to show a gradual crossover
from having the characteristic of delocalized states, $r_{{\rm ext}}\approx0.6$,  when $\xi \gg L$, to the value that indicates localized behavior, $r_{{\rm ext}}\approx 0.39$, when $\xi \ll L$.

This behavior is demonstrated in Figs.~\ref{fig:smalldisordertrans}(a), panels (i)--(iii), where we plot the disorder averaged level spacing ratio $r$ and the density of Floquet states, as a function of the quasi-energy for different disorder strengths. For weak disorder, $VT=0.5$, panel (i) shows that the level spacing ratio is $r \approx 0.6$ in any spectral region where
Floquet states exists. On the other hand, panel (iii) shows that
already for $VT=4$, the level-spacing ratio approaches $r \approx 0.39$
at all quasi energies, as expected from localized states.

Note that, as the disorder strength increases, the level spacing ratio decreases uniformly throughout the spectrum [Fig.~\ref{fig:smalldisordertrans}(a), panels (i--iii); the same behavior is seen at weaker values of the disorder (not shown)]. There is no quasi-energy in which the LSR remains close to $0.6$, corresponding delocalized Floquet eigenstates. This is consistent with the expectation that the bulk Floquet states become localized even for weak disorder, and the localization length becomes shorter as the disorder strength increases. The behavior of the LSR as a function of system size, Fig.~\ref{fig:smalldisordertrans}(b), also shows behaviour consistent with the above expectation.
In contrast, if the bulk bands of the clean systems carried non-zero Chern numbers, delocalized states would persist in the bands up to a critical strength of the disorder, at which point they would merge and annihilate.


%

In the AFAI phase all the bulk states are localized, but the edge hosts chiral modes at any quasi-energy (cf.~Sec.~\ref{sec: top invariant}). 
To test this, we simulate 
the time evolution of wavepackets initialized either in the bulk or near the edge of the system. We consider the system in a rectangular geometry.  The initial state,
$|\psi_{0}\rangle$, is localized to a single site $\mathbf{x}_{0}=\left(x_{0},y_{0}\right)$. 
To obtain information on quasi-energy resolved propagation, we investigate the disorder-averaged transmission probability, $\overline{ \big|G_{N}\left({\bf x},{\bf x}_{0},\ve\right)\big|^2 }$, which is a function both of quasi-energy $\ve$ and the total time of evolution $T_{f}=NT$. Here, the bar denotes disorder averaging. The transmission amplitude in each disorder realization, $G_{N}$, is
obtained by a partial Fourier transform of the real time amplitude, $\tilde{G}\left({\bf x},{\bf x}_{0},t\right) = \left\langle {\bf x}\left|U(t)\right|\psi_{0}\right\rangle$, and is given by
\begin{equation}
G_{N}\left({\bf x},{\bf x}_{0},\ve\right)  =\frac{1}{N} \sum_{n=0}^{N}\tilde{G}\left({\bf x},{\bf x}_{0},t=nT\right)e^{i\ve nT}.
\label{eq: GN}
\end{equation}

\begin{figure}

\includegraphics[width=1\columnwidth]{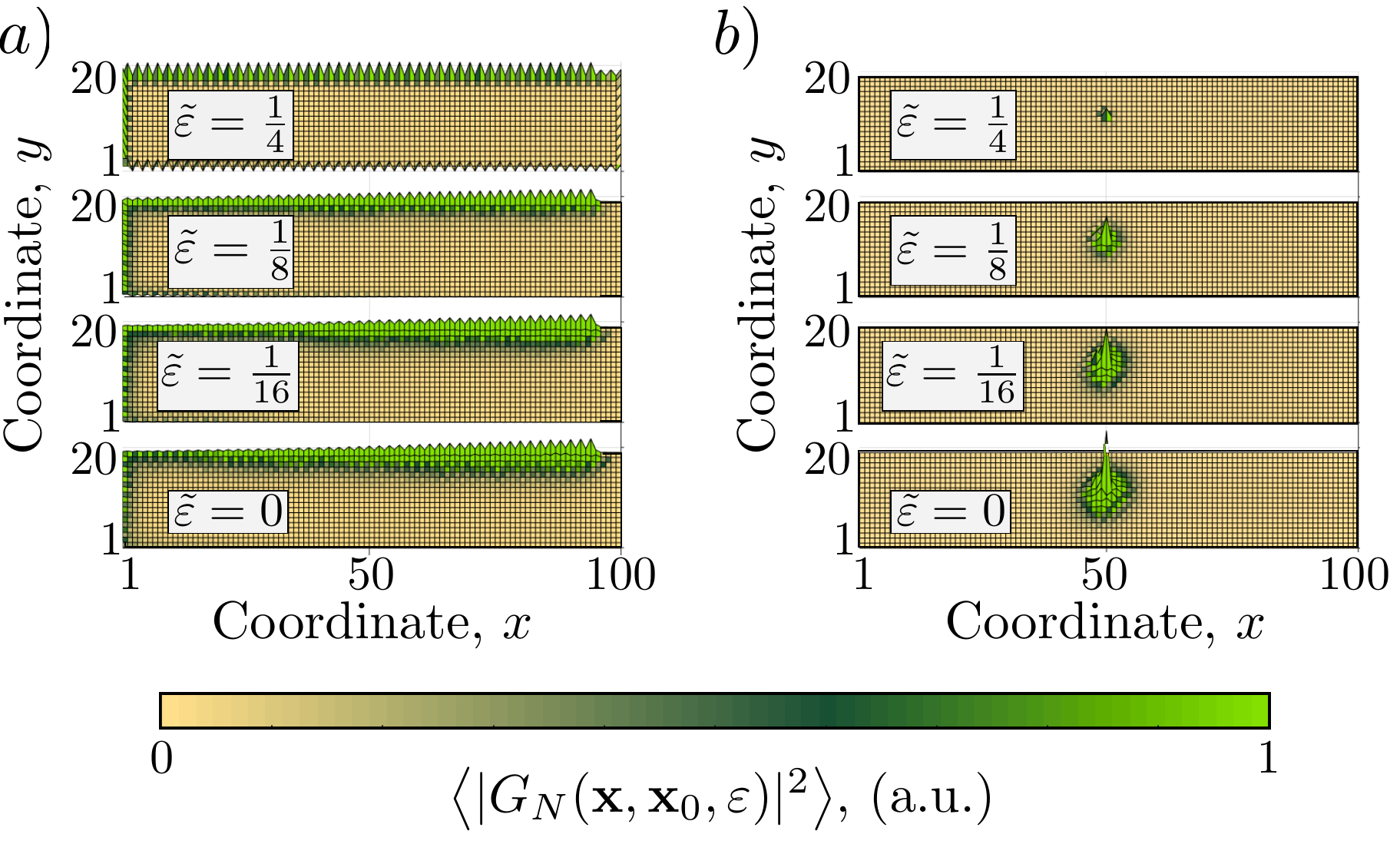}
\caption{
Wavepacket dynamics in the AFAI. Using the same model as in Fig.~\ref{fig:smalldisordertrans}, we plot the amplitude of the transmission probability,$\avg{|G_{N}\left({\bf x},{\bf x}_{0},\ve\right)|^2}$, c.f. Eq.~\eqref{eq: GN}
obtained after a time-evolution of $T_{{\rm fin}}=300T$ and averaged over disorder realizations. We simulate a strip of size $20\times 100$ with open boundary conditions, and plot $\avg{|G_{N}|^2}$ for several
quasienergies $\epsilon/\Omega=0$, $\frac{1}{16}$, $\frac{1}{8}$,
$\frac{1}{4}$. (a) shows $\avg{|G_{N}|^2}$ when the initial wavepacket
is chosen at the edge $\mathbf{x}_0=\left(96,1\right)$. It indicates
the presence of a robust edge mode at all the given quasi-energies.
(b) shows the probability when the initial wavepacket is chosen in
the bulk, $\mathbf{x}=\left(50,10\right)$. This indicates that the
bulk Floquet states are localized. These simulations were carried out
with a time step of $dt=T/100$.}\label{fig:timeevolv}
\end{figure}
The real time transmission amplitude $\tilde{G}(t)$ is computed numerically by a split operator
decomposition. Figs.~\ref{fig:timeevolv}(a),(b) show 
$\overline{|G_{N}|^2}$ at different
quasi-energies, for initial states on the edge and in the bulk, respectively. The simulations are done for a disorder strength
$VT=4$. At this disorder strength, the analysis of the level-spacing statistics shown in Fig.~\ref{fig:smalldisordertrans}(a) indicates that all the bulk Floquet bands
are localized with a localization length smaller than the system size.
Fig.~\ref{fig:timeevolv}(a) shows the value of $\overline{|G_{N}|^2}$ when
the wavepacket is initialized at the edge of the system, $\mathbf{x}_0 = (1,1)$. The wave packet propagates chirally along the edge. The figure exemplifies that the edge modes are robust in the presence of disorder, and are present at all quasi-energies.
Importantly, edge states are also
observed at quasi-energies  where the bulk density of states is appreciable, indicating that the chiral edge states coexist with localized bulk states [the density of states in the bulk is shown in Fig.~\ref{fig:smalldisordertrans}(a)].

In contrast, Fig.~\ref{fig:timeevolv}(b) shows $\overline{|G_N|^2}$ for a wavepacket initialized in the middle of the system. The wavepacket remains localized at all quasi-energies, as expected if all bulk Floquet eigenstates are localized.
This confirms that the model we study numerically indeed exhibits the basic properties of the AFAI phase: fully localized Floquet bulk states, coexisting with chiral edge states which exist at every quasi-energy.

\begin{figure}
\includegraphics[width=1.0\columnwidth]{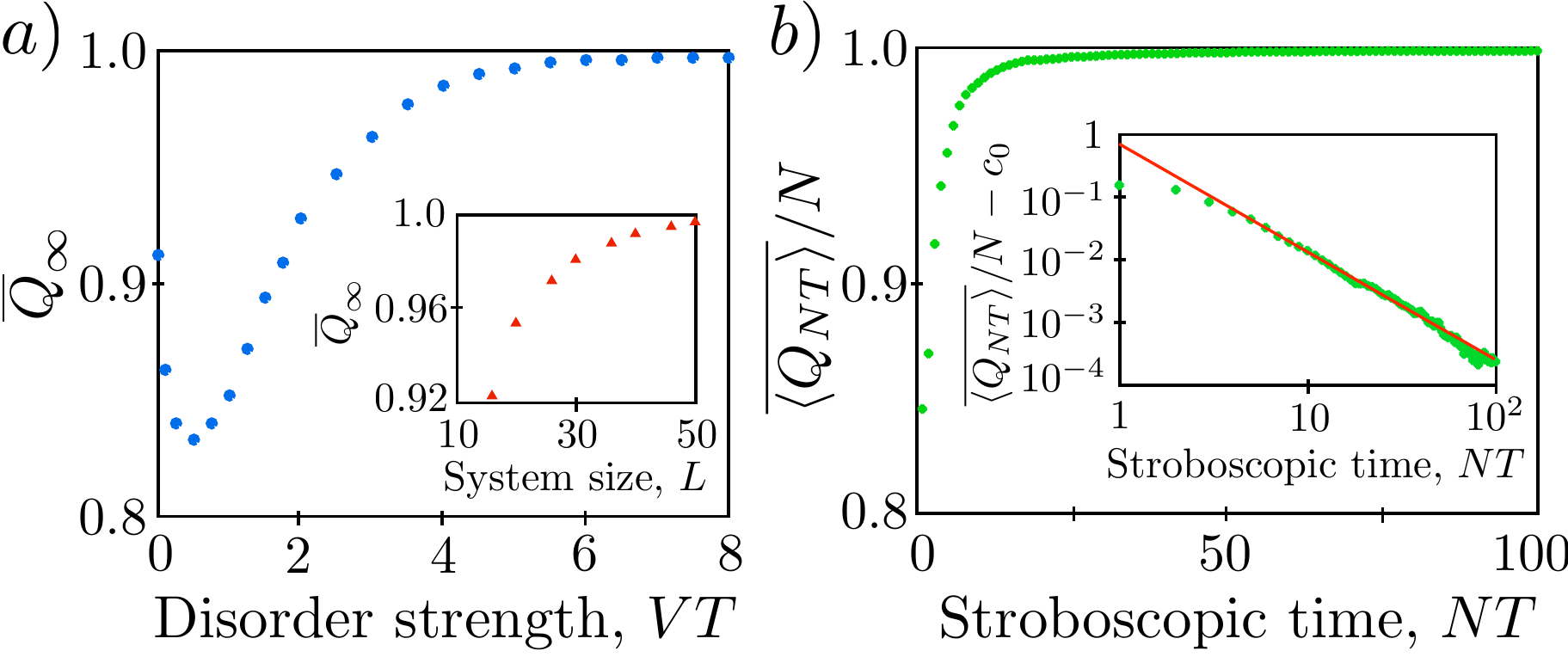} 
\caption{Quantized charge pumping in the AFAI. (a) Cumulative average of the
pumped charge per cycle in the limit of long times, $\overline{Q}_\infty$, [c.f. Eq.~(\ref{eq: Q infty})], as a function of disorder strength. For $VT\gtrsim 5$, the localization length is sufficiently smaller than the system size, and $\overline{Q}_\infty$ approaches unity. The inset shows the finite size scaling of $\overline{Q}_\infty$ for $VT=8$. (b) Cumulative average of the pumped charge for $N$ periods, $\overline{\avg{Q}}_{NT}/N$, as a function of $N$.  The disorder strength used was $VT=8$. The approach to the quantized value can be fit to a power law $(NT)^{-\upsilon}$ with $\upsilon=1.72$, see the log-log plot shown in the inset. In both panels, we averaged the  charge pumped across all the lines running parallel to the $y$ direction  of the cylinder (see Fig.~\ref{fig: cylinder}) and over 100 disorder realizations. The system size used $L_x \times L_y=50\times50$.}  \label{fig:chargepumping}
\end{figure}

Next, we numerically demonstrate the quantized charge pumping property of the AFAI. Using the model described above, we numerically compute the value of $\overline{Q}_\infty$ given by  Eq.~\eqref{eq: Q infty} for a single value of the flux, $\theta_x=0$. When computing $\overline{Q}_\infty$, we averaged the charge pumped across all the lines running parallel to the $y$ direction of the cylidner (see Fig.~\ref{fig: cylinder}), as well as over 100 disorder realizations. 
In Fig.~\ref{fig:chargepumping}(a), we show the cumulative average of the
pumped charge per cycle in the limit of long times, $\overline{Q}_\infty$ [c.f. Eq.~\eqref{eq: Q infty}]  as a function of disorder strength. At weak disorder, when the localization length is smaller than the system size, $\overline{Q}_\infty$ is clearly not quantized. However, for disorder strength $VT\gtrsim 5$, the value of $\overline{Q}_\infty$ quickly tends towards unity.   This agrees with the results presented in Fig.~\ref{fig:smalldisordertrans}(a.iii), which indicate that at this disorder strength, the localization length is substantially smaller than $L=70$. Finite size scaling demonstrating that  $\overline{Q}_\infty$ indeed asymptotes to unity in the thermodynamic limit is presented in the inset of  Fig.~\ref{fig:chargepumping}(a).

The value of the cumulative average of the pumped charge  over $N$ periods, $\overline{\avg{Q}}_{NT}/N$ [c.f. Eq.~\eqref{eq: current def}] is plotted vs.~$N$ in Fig~\ref{fig:chargepumping}(b), demonstrating its approach to $\overline{Q}_\infty$ for large values of $N$ (i.e., at long times). As in panel (a), we averaged over all the lines running parallel to the $y$ direction, and over 100 disorder realizations.   We examine the asymptotic behavior of $\overline{\avg{Q}}_{NT}$ 
and find a power law behaviour of the form $\overline{\avg{Q}}_{NT}=\overline{Q}_\infty+c N^{-\upsilon}$ with $\upsilon=1.72$, shown in the inset of panel (b). Note that for a \textit{single} disorder realization and a \textit{single} vertical cut, $\avg{Q}_{NT}$ is expected to exhibit an oscillatory behaviour with an envelope which decays as $1/N$, see Appendix~\ref{appendix: spectral flow}. This expectation is indeed confirmed by our numerical simulations, as we show in Appendix~\ref{appendix: single pump}. In contrast, Fig.~\ref{fig:chargepumping}(b) shows a power-law behaviour with a power larger than $1$ and no oscillations; this is clearly the result of averaging over the frequencies appearing in  $\avg{Q}_{NT}/N$ for each disorder realization and vertical cut. The above results numerically confirm the discussion in Sec.~\ref{sec: charge pumping}, and conclude our numerical analysis of the AFAI phase.


\subsection{Strong Disorder Transition}
\label{sec: strong disorder}
For sufficiently strong disorder, we expect the AFAI to give way to a topologically trivial localized phase in which the winding number vanishes.
We now analyze the transition between the AFAI and this ``trivial'' phase. 
As explained above, the winding number $W_\ve$ can only change if a delocalized state crosses through the quasi-energy $\ve$ as disorder is added.  In the AFAI phase all of the bulk states are already localized. How does the transition between the two phases occur?

Clearly, 
at the transition, delocalized states must appear in the quasi-energy spectrum. As disorder is increased, the delocalized states must sweep the full quasi-energy zone, changing the topological invariant $W_\ve$ as they do so. The transition from the AFAI phase to the trivial phase can therefore occur through a range of disorder strength $V^-_{c}<V<V^+_{c}$, where $V^-_{c}$  is the disorder strength at which the first delocalized state appears, and $V^+_{c}$ is the disorder strength at which all Floquet states are again localized, and $W_\ve=0$ for all $\ve$. Below we will support this scenario using numerical simulations, and furthermore provide evidence suggesting that the transition is of the quantum Hall universality class.



\begin{figure}
\includegraphics[width=1.0\linewidth]{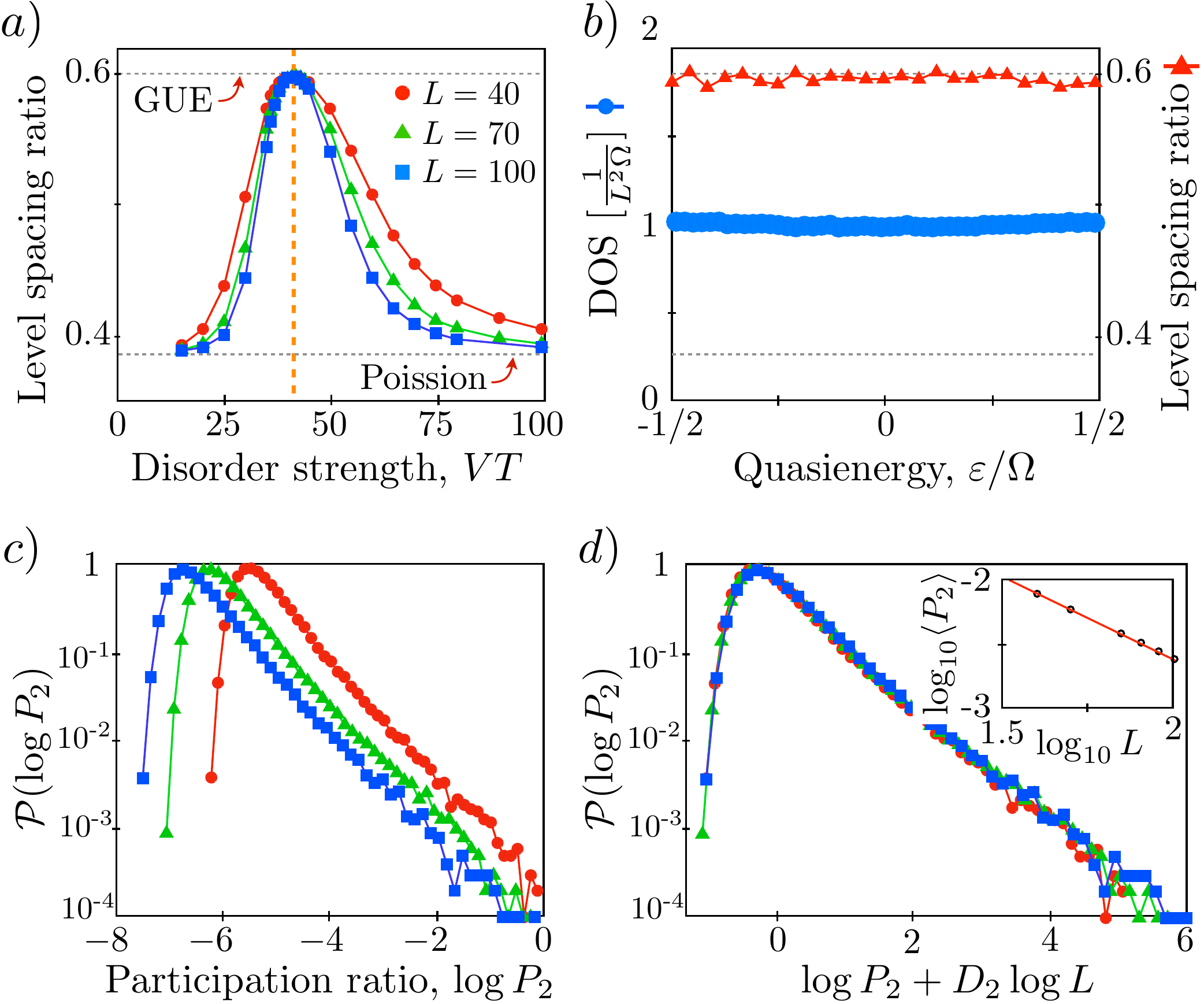}
\caption{Transition from the AFAI into a trivial phase at strong disorder.
(a) Average level spacing ratio as a function of disorder
strength. On increasing disorder strength, a transition is observed
between two localized phases with delocalized levels at $VT\approx40$.
Here, the level-spacing ratio has been averaged over all quasi energies.
(b) Level spacing ratio as a function of quasienergy and
its comparison with the DOS, indicating that the entire Floquet
band is delocalized. (c) Effect of finite size of on the distribution
of the participation of ratio, $P_{2}$, at a given disorder strength,
$VT=40$. The system sizes used for the simulations are $L_{x}\times L_{y}=40\times40$,
$70\times70$, $100\times100$. The shape of the curve does not change,
indicating a critical phase. (d) Scaling collapse of the
three curves with $D_{2}=1.3$, where for a critical phase it is expected
that $\left\langle P_{2}\right\rangle \sim L^{-D_{2}}$.\label{fig:strongdisordertrans}}
\end{figure}
We study the same model used in Sec.~\ref{sec: numerics weak} and examine the level-spacing ratio, $r$, as a function of disorder
strength and quasi-energy. For this model, our simulations indicate $V^-_{c}\approx V^+_{c}$, within our resolution 
(limited by the system size).  
In Fig.~\ref{fig:strongdisordertrans}(a), we plot $r$, averaged over disorder realizations
and all quasi-energies. We see that at disorder strength $V_cT\approx40$
the level spacing ratio reaches $r\approx0.6$, indicating delocalization. 
On either side of this point, $r$ approaches $0.39$ as the system size increases, which indicates localization. 
The peak in the value of $r$ as a function of disorder
gets sharper for larger system size, which is a signature of
a critical point of this transition. In Fig~\ref{fig:strongdisordertrans}(b), we show that at disorder strength $V_{c}$, the LSR is independent of the quasi-energy with $r\approx 0.6$ (for disorder strengths close to $V_c$, we also find that the LSR is independent of the quasi-energy, but with $r<0.6$). This indicates that all of the Floquet states have a delocalized character at this disorder strength, which leads us to conclude that $V_{c}=V^-_c\approx V^+_{c}$.


At the critical point, $V=V_{c}$, we expect the wavefunctions
to have a fractal character~\cite{Ludwig1994}. This behavior is manifested in the distribution of the inverse participation
ratio (IPR), $P_{2}=\sum_\br\left|\psi(\br)\right|^{4}$. 
We study the distribution of the IPR, $\mathcal{P}\left(\log P_{2}\right)$, among all the Floquet eigenstates and averaged over disorder realizations. Fig.~\ref{fig:strongdisordertrans}(c) shows
the distribution for different system sizes.
We note that the shapes of the distributions for different sizes
are similar, a signature of criticality. In two dimensions, the average value of the IPR at a critical point is expected
to scale like $\left\langle P_{2}\right\rangle \sim L^{-D_{2}}$,
with $D_{2}<2$ \cite{Ludwig1994}. Fig.~\ref{fig:strongdisordertrans}(d) shows the
scaling collapse of all the distributions. From the collapse we find the
fractal dimension, $D_{2}=1.3$. The inset in this figure also
shows a linear scaling $\log\left\langle P_{2}\right\rangle \sim-D_{2}\log L$.
The critical exponent $D_2$ we find in our numerical simulations is close to the value found for the universality class of quantum Hall plateau transitions~\cite{Ludwig1994,Huckestein1992}, $D_2 \approx 1.4$, indicating that the transition from the AFAI to the trivial phase may belong to this universality class. This is natural to expect, since, like the quantum Hall transition, in the transition out of the AFAI phase a delocalized state with a non-zero Chern number must ``sweep'' through every quasi-energy, to erase the chiral edge states.  We expect that the AFAI transition can be described in terms of “quantum percolation” in a disordered network model, similar to the Chalker-Coddington model for the plateau transitions~\cite{Chalker1988}. We leave such investigations for future work.

\section{Discussion}

In this paper we have demonstrated the existence of a new non-equilibrium phase of matter: the anomalous Floquet-Anderson insulator.
The phase emerges in the presence of time-periodic driving and disorder in a two-dimensional system, and features a unique combination of chiral edge states and a fully localized bulk.
Such a situation cannot occur in non-driven systems, where the presence of chiral edge states necessarily implies the existence of delocalized bulk states where the chiral branches of the spectrum can terminate.
In a driven system, the periodicity of the quasienergy spectrum alleviates this constraint, allowing chiral states to ``wrap around'' the quasienergy zone and close on themselves.

One of the key physical manifestations of the AFAI is a new type of {\it non-adiabatic} quantized pumping, which occurs when all states near one edge of the system are filled.
It is interesting to compare this phenomenon with Thouless' quantized {\it adiabatic} pumping, described in Ref.~\cite{Thouless1983}.

\begin{figure}
\includegraphics[width=1.0\linewidth]{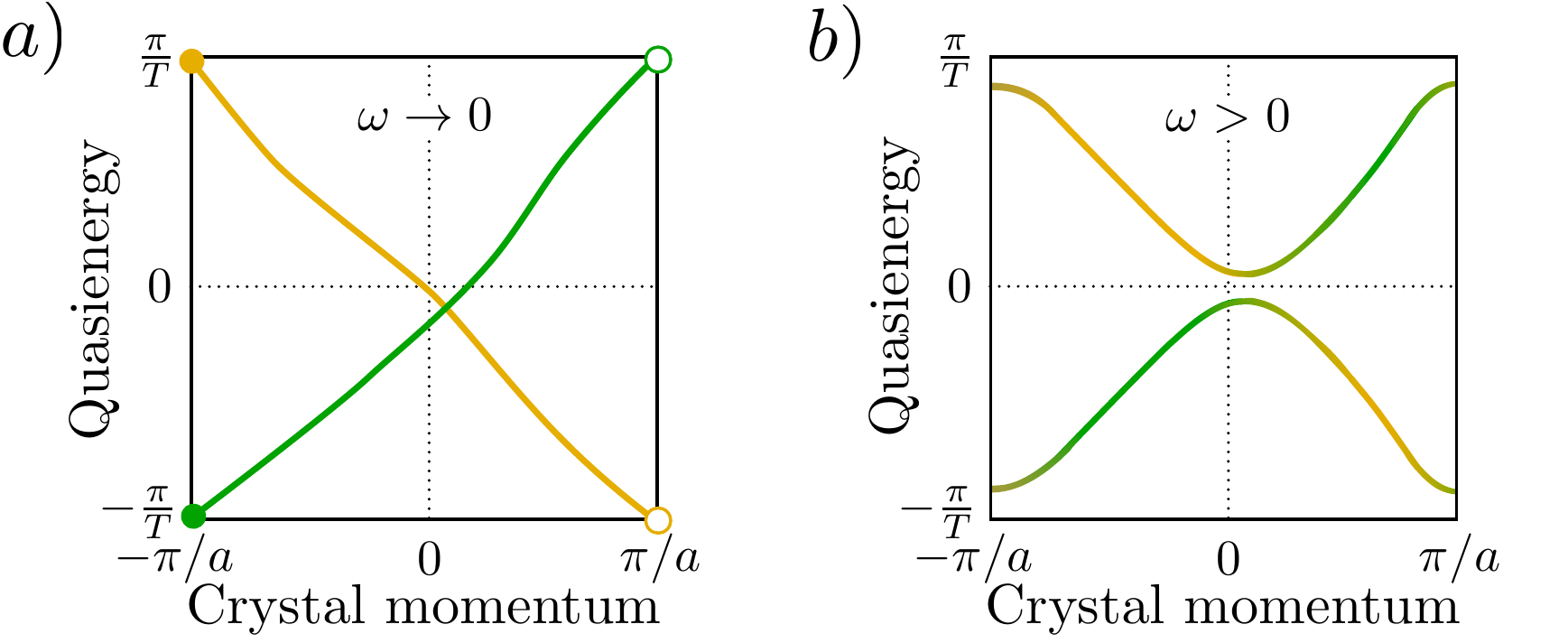}
\caption{
\label{fig:Thouless}
Floquet spectrum for Thouless' quantized adiabatic charge pump.
a) Quantized adiabatic pumping in a 1D system is manifested in chiral Floquet bands that wind around the quasienergy Brillouin zone (right and left movers are shown in green and orange, respectively).
b) Outside of the adiabatic limit, $\omega > 0$, counterpropagating states hybridize, and all Floquet bands obtain trivial winding numbers; quantized pumping is destroyed.
}
\end{figure}
The complementary relationship between pumping in the AFAI and the Thouless case is best revealed by first viewing the Thouless pump from the point of view of its Floquet spectrum.
In Thouless' one-dimensional pump, a periodic potential is deformed adiabatically such that in each time cycle a quantized amount of charge is pumped through the system. In the adiabatic limit, the quasi-energy spectrum of the pump exhibits one pair of counter propagating one-dimensional chiral Floquet-Bloch bands, which wrap around the quasi-energy Brillouin [see Fig.~\ref{fig:Thouless}(a)].
The (nonzero) quasi-energy winding number of each band gives the associated quantized pumped charge~\cite{KBRD}.
Importantly, for any finite cycle time the two counter-propagating states hybridize and destroy the perfect quantization of the charge pumped per cycle [Fig.~\ref{fig:Thouless}(b)].
%

In a strip geometry, the AFAI can be viewed as a quasi-one-dimensional system.
As discussed in Sec.~\ref{sec: top invariant}, the system hosts chiral edge states that run in opposite directions on opposite edges.
Furthermore, as shown by the spectral flow (see Fig.~\ref{fig: cylinder}), these counterpropagating chiral modes cover the entire quasi-energy zone, analogous to the counter-propagating modes of the Thouless pump [Fig.~\ref{fig:Thouless}(a)].
Crucially, however, the counterpropagating modes of the AFAI are {\it spatially separated} and therefore their coupling is {\it exponentially suppressed}: no adiabaticity restriction is needed to protect quantization.
Thus quantized pumping at finite frequency can be achieved in the AFAI phase.

How is the AFAI manifested in experiments?
First, the localized bulk and chiral propagating edge states could be directly imaged, for example in cold atomic or optical setups.
More naturally for a solid state electronic system, the pumping current could be monitored in a two terminal setup.
Unlike the case of an adiabatic pump, where a quantized charge is pumped at zero source-drain bias, to observe quantized charge pumping in the AFAI the chiral propagating states of one edge of the system would need to be completely filled at one end of the sample, and emptied at the opposite end.
We speculate that this can be achieved using a large source-drain bias.
A detailed analysis of such non-equilibrium transport in a two or multi-terminal setup, as well as an investigation of promising candidate systems, are important directions for future study.

The implications of our results go beyond those specific to the class of systems studied in this paper.
As a direct generalization of our results, one can consider constructing anomalous Floquet insulators in different dimensions and symmetry classes.  Floquet-Bloch band structures which generalize those of Ref.~\cite{Rudner2013} can serve as a starting point for constructing such anomalous periodically driven systems.
Going beyond the single particle level, an important challenge is to understand how the properties of the AFAI change in the presence of interactions.
An exciting possibility is to obtain a topologically non-trivial steady state for an interacting, periodically driven system \cite{Dehghani2014, Dehghani2014b, Bilitewski2014, Seetharam2015, Iadecola2015,SondhiUnpublished}. The common wisdom dictates that a periodically system with dispersive modes is doomed to evolve into a highly random state which is essentially an infinite temperature state as far as any finite order correlation functions are concerned \cite{DAlessio2014,Ponte2014,Ponte2014b,LazaridesDasMoessner2014}.
Our results on the single particle level demonstrate that it is possible to obtain a topological Floquet spectrum with no delocalized states away from the edges of the system.
It is therefore possible that such periodically driven systems can serve as a good starting point for constructing topologically non-trivial steady states for interacting, disorder (many-body localized) periodically driven systems.
What types of topological steady states can be obtained by this method, and what are their observable signatures, 
will be interesting subjects for future work.

\begin{acknowledgments}
We are grateful to Cosma Fulga, Mykola Maksimenko, and Ady Stern for enlightening discussions. P.T. and N.L. acknowledge support from the US - Israel Bi-National Science foundation. N.L acknowledges support from the CIG Marie Curie grant and from and I-Core, the Israeli
excellence center “Circle of Light”.
E. B. was supported by the Minerva foundation, the CIG Marie Curie grant, and the Israel Science foundation.
M. R. acknowledges support from the Villum Foundation and from the People Programme (Marie Curie Actions) of the European Union's Seventh Framework Programme (FP7/2007-2013) under REA grant agreement PIIF-GA-2013-627838. G. R.  and P.T. are grateful
for support from NSF through DMR-1410435, as well as the Institute of Quantum Information and matter, an
NSF Frontier center funded by the Gordon and Betty Moore Foundation, and the Packard Foundation.
\end{acknowledgments}

\appendix
\section{Construction of the deformed evolution operator on the cylinder}
\label{app:flattening}
In this appendix we construct the deformed evolution operator on the cylinder, $\mathcal{\tilde{U}}_\ve(t)$ of Eq.~(\ref{eq: U matrix}), which is used to demonstrate the existence of delocalized edge states in the AFAI.  The deformation is designed such that at $t = T$ it interpolates smoothly between $\tilde{U}(T)$ (corresponding to the \textit{cylinder}) in the vicinity of the edges of the cylinder, and $\mathds{1}$ in the bulk of the cylinder. We first define the family of operators $\cF(s) = \sum_\br \alpha(y,s)|\br\rangle\langle \br |$, where
\begin{equation}
\alpha(y,s)=\begin{cases}
0, & y\le\ell_{1}\\
s \frac{(y-\ell_1)}{(\ell_2-\ell_{1})}, & \ell_{1}\le y\le\ell_2\\
s, & \ell_2<y<L_{y}-\ell_{2}\\
s\frac{(L_{y}-\ell_1-y)}{(\ell_2-\ell_{1})} , & L_y-\ell_2 \le y \le L_y-\ell_{1}\\
0, & y\ge L_{y}-\ell_{1}.
\end{cases}
\end{equation}
Here, we choose $\xi \ll \ell_1 \ll \ell_2 \ll \ell_0$, where $\xi$ is the bulk localization length. Analogously to Eq.~(\ref{eq: def U epsilon}), the family of deformed evolution operators is defined as
\begin{equation}
\mathcal{\tilde{U}}_\ve(t,s) = U(t) \exp\left[ i t \cF(s)  H_\ve^{\mathrm{eff}} \cF(s) \right].
\label{eq:flattenedU}
\end{equation}
Here, $H_\ve^{\mathrm{eff}}$ is defined as in Eq.~(\ref{eq: def U epsilon}), i.e., $H_\ve^{\mathrm{eff}}=\frac{i}{T}\log U(T)$, where $U(T)$ is the evolution operator for a full period \emph{on the torus}.
The deformed evolution operator corresponds to $\mathcal{\tilde{U}}_\ve(t) \equiv \mathcal{\tilde{U}}_\ve(t, s=1)$.



Note that, strictly speaking, $\mathcal{\tilde{U}}_\ve(t)$ of Eq.~(\ref{eq:flattenedU}) is not precisely of the block-diagonal form of Eq.~(\ref{eq: U matrix}). It still has exponentially small but non-zero matrix elements connecting the different blocks. However, a second deformation can take $\mathcal{\tilde{U}}_\ve(T)$ to the form appearing in Eq.~\eqref{eq: U matrix}.

\section{Perturbative derivation of the effective Hamiltonian}
\label{appendix: perturbation}

Here, we outline the details of the derivation used to demonstrate the perturbative stability of the AFAI phase. We first examine the effective static Hamiltonian defined in Eq.~(\ref{eq: static heff}), expressed as a power series in $\lambda$. We insert Eq.~(\ref{eq:Deff}) into (\ref{eq: static heff}) and expand both sides in powers of $\lambda$. For the left-hand side, we obtain
\begin{align}
\label{eq:Ueff}
& e^{-iT\left(H_{\mathrm{eff}}^{0} + D_{\mathrm{eff}}\right)}  \\
& =U^{0}_T\left[1-i\int_{0}^{T}dt\left(\lambda D_{\mathrm{eff}}^{(1)}(t)+\lambda^{2}D_{\mathrm{eff}}^{(2)}(t)+\dots\right)\right.\nonumber \\
 & -\int_{0}^{T}\!\!\!\!\int_{0}^{t}dtdt'\left(\lambda D_{\mathrm{eff}}^{(1)}(t)+\dots\!\!\right)\left(\lambda D_{\mathrm{eff}}^{(1)}(t')+\dots\!\!\right)\left.+\vphantom{\int}\dots\right], \nonumber
\end{align}
where $U^0_t=e^{-it H_{\mathrm{eff}}^{0}}$ and $D_{\mathrm{eff}}^{(n)}(t)=(U^0_t)^\dagger D_{\mathrm{eff}}^{(n)}U^0_t$.
Note that the eigenvalues of $H_{\mathrm{eff}}^{0}$ are only defined modulo $2\pi k_{j}/T$ (where $k_j$ is an integer).
The form of $D_{\mathrm{eff}}^{(n)}$(t) depends on the choice of $k_j$, while the evolution operator does not.
To fix this ambiguity, we choose $E_j$ to lie in the range $[-\pi/T, \pi/T)$.


The right--hand side of~(\ref{eq: static heff}), expanded in powers of $\lambda$, reads
\begin{align}
U(T) & =U_{0}(T,0)\left[1-i\lambda\int_{0}^{T}dt\cD(t)\right.\nonumber\\
&-\left.\lambda^{2}\int_{0}^{T}dt\int_{0}^{t}dt'\cD(t)\cD(t')+\dots\right].\label{eq:U}
\end{align}
where $U_0(t,t') = \mathcal{T} \mathrm{exp} \left[-i\int_{t'}^{t} H_0(t) \right]$ and $\cD(t)=U_{0}(0,t) D(t) U_{0}(t,0)$.


Equating (\ref{eq:Ueff}) and (\ref{eq:U}), and using $e^{-iTH_{\mathrm{eff}}^{0}}=U_0(T,0)$,
we find that
\begin{equation}
\int_{0}^{T}dt D_{\mathrm{eff}}^{(1)}(t)=\int_{0}^{T}dt \cD(t),\label{eq:W1}
\end{equation}
and likewise
\begin{align}
\int_{0}^{T}dt D_{\mathrm{eff}}^{(2)}(t) & = -i \int_{0}^{T}dt\int_{0}^{t}dt' \cD(t) \cD(t')\nonumber \\
 & +i\int_{0}^{T}dt\int_{0}^{t}dt'D_{\mathrm{eff}}^{(1)}(t)D_{\mathrm{eff}}^{(1)}(t'),
 \label{eq:W2}
\end{align}
and so forth.

To find $D^{(n)}_{\mathrm{eff}}$ explicitly, we express them in the ``tight-binding'' form~(\ref{eq:tightbind}); inserting this form into~(\ref{eq:W1},\ref{eq:W2}), and using the fact that $H^0_{\mathrm{eff}}$ contains only on-site potentials and no inter-site hopping, we arrive at
\begin{align}
\int_0^T dt D_{\mathrm{eff}}^{(n)}(t) &= \int_0^T dt \sum_{i,j} e^{i E_{ij} t} \Delta^{(n)}_{i,j} c^\dagger_i c^{\vphantom{\dagger}}_j \nonumber \\
&= \sum_{i,j} \frac{e^{i E_{ij} T} - 1}{i E_{ij}} \Delta_{i,j}^{(n)} c^\dagger_i c^{\vphantom{\dagger}}_j.
\label{eq:Dn}
\end{align}
Equating this expression for $n=1$ to the right hand side of Eq.~(\ref{eq:W1}) gives Eq.~(\ref{eq: delta0}). From~(\ref{eq:Dn}) we see that, as long as $E_{ij} T<2\pi$ for every pair of sites, $\Delta^{(n)}_{i,j}$ is non-singular. As $E_{ij} T \rightarrow 2\pi$, $\Delta^{(n)}_{i,j}$ may diverge for all $n$, and the expansion in $\lambda$ fails. This reflects the fact that, in general, a Floquet operator whose eigenvalues are spread throughout the quasi-energy zone cannot be generated by a static, \emph{local} Hamiltonian.

From the form of the right-hand side of Eq.~(\ref{eq:W1}), we can analyze the maximum range of the hopping matrix elements in $D_{\mathrm{eff}}^{(1)}$. We denote the maximum range of the hopping matrix elements in $D(t)$ by $r$, where $r=1$ corresponds to a nearest neighbor hop, $r=2$ to second neighbors, and so on. Since the matrix elements of the unperturbed evolution operator $U_0(t,0)$ vanish beyond second neighbor sites on the square lattice, we find that $D_{\mathrm{eff}}^{(1)}$ contains matrix elements whose range is at most $r+4$.
Similarly, from Eq.~(\ref{eq:W2}), $\Delta^{(2)}_{ij}$ vanishes beyond the $2r+6$th neighbor, and more generally, $\Delta_{i,j}^{(n)}$ vanishes beyond range $n(r+2)+2$. Hence, the matrix elements of $D_{\mathrm{eff}}$ at range $n(r+2)+2$ contain the exponentially small factor $\lambda^n$.

\section{Quantized charge pumping and the winding number}
\label{appendix: current}
In this Appendix, we show that for the AFAI, a non-zero value for the winding number $W_\ve$ implies quantized charge pumping on the edge of the system. As in Sec~\ref{sec: charge pumping}, we take an initial state with all {\it sites} filled in a strip of width $\ell$ near one edge of the AFAI, [see Fig.~\ref{fig: cylinder}(c)], and the rest to be empty.

To calculate the time-dependence of the pumped charge, we begin by deriving an expression for the instantaneous current flowing across a longitudinal cut through the cylinder, i.e., across a line parallel to the $y$-axis. 
The corresponding current operator is found by first allowing a flux $\theta_x$ to be threaded through the cylinder.
Next we pick a gauge where the gauge (vector) potential is nonzero only on the links connecting sites with $x = L_x$ to sites with $x = 0$.
The net current flowing across the cut between $x = L_x$ and $x = 0$ is then described by the operator $I_x(t) = \partial\tilde{H}(\theta_x,t)/\partial\theta_x$, where $\tilde{H}(\theta_x,t)$ is the Hamiltonian of the system in the presence of the flux $\theta_x$. Here the tilde denotes the cylindrical geometry.

Below, we first (Sec.~\ref{appendix: spectral flow})  show that when averaged over many periods, the charge pumped approaches a quantized value $Q_\infty$  equal to the edge topological invariant $n_{\mathrm{edge}}$, expressed in terms of the spectral flow on one edge of the system [Eq.~(\ref{eq: current spec flow})]. We then show (Sec.~\ref{appendix: winding}) that $Q_\infty$ is in fact equal to the bulk topological invariant $W_\ve$, given by Eq.~\eqref{eq: invariant} \\

\subsection{Quantized charge pumping from spectral flow}
\label{appendix: spectral flow}
We start from Eq.\eqref{eq: current def}, which gives the charge pumped during the time integral $0<t<\tau$.  The initial state, which is a single Slater determinant in terms of position eigenstates, is given by a superposition of Slater determinants in term of Floquet states
\begin{equation}
|\Psi\rangle=\sum_{\cS}A_\cS\prod_{j\in\cS}\psi^\dagger_j|0\rangle.
\end{equation}
Inserting this into the expression for the pumped charge in the interval $0<t<\tau$, Eq.~\eqref{eq: current def}, we get
\begin{equation}
\langle Q\rangle _\tau = \sum_{\cS,\cS'}A^*_\cS A_{\cS'}\sum_{j \in \cS}\sum_{k \in \cS'}  \int_0^T dt \langle \psi_j(t)| \frac{\partial \tilde{H}(\theta_x,t) }{\partial \theta_x}|\psi_k(t)\rangle.
\label{eq: current full}
\end{equation}
The double sum in Eq.~\eqref{eq: current full} contains both diagonal and off diagonal terms for the contributions of single particle Floquet states. Denoting each of the contributions by $Q_{jk}$, and using $i\partial_t |\psi(t)\rangle=\tH(t)|\psi(t)\rangle$, the different terms in (\ref{eq: current full}) can be written as
\begin{align}
Q_{jk}  &=  \int_0^T dt \langle \psi_j(t) | \left\{ \partial_{\theta_x} (\tilde{H} | \psi_k(t) \rangle) - \tH \partial_{\theta_x}| \psi_k(t)\rangle \right\} \nonumber\\
&=\int_0^T dt\Big\{ \langle \psi_j(t)|\partial_{\theta_x}i\partial_t|\psi_j(t)\rangle\nonumber\\
&\qquad\qquad\qquad+i\partial_t\left(\langle\psi_j(t)|\right) \partial_{\theta_x}|\psi_k(t)\rangle\Big\}\nonumber\\
&= i\int_0^T dt \partial_t \langle \psi_j(t)|\partial_{\theta_x}|\psi_k(t)\rangle.
\label{eq: current 2}
\end{align}

According to Floquet's theorem, the Floquet states  can be written as $|\psi_j(t)\rangle=e^{-i\ve_j t}|\phi(t)\rangle$, where $|\phi_j(t)\rangle=|\phi_j(t+T)\rangle$ is a periodic function. Substituting this into Eq.~(\ref{eq: current 2}) we obtain for the diagonal terms,
\begin{align}
Q_{jj}  &=  T\frac{\partial{\ve_j}}{{\partial\theta_x}},
\end{align}
and for the off diagonal terms,
\begin{align}
Q_{jk}  &= i\left(e^{i\left(\ve_j-\ve_k\right)T}-1\right) \langle\phi_j(0)|\partial_{\theta_x}|\phi_k(0)\rangle, \;\; j\neq k.
\end{align}
Clearly, when computing the average charge pumped over $N$ periods, $\avg{Q}_{NT}/N$, the off-diagonal terms will give a contribution which decays as $1/N$, while the diagonal terms will give the contribution which does not decay with $N$,
\begin{equation}
\lim_{N\to\infty}\avg{Q}_{NT}/N=Q_\infty= T \sum_j n_j \frac{\partial{\ve_j}}{{\partial\theta_x}},
\label{eq: charge no theta avg}
\end{equation}
where $n_j$ is the probability for the $j^{\mathrm{th}}$ Floquet state to be occupied, $n_j=\langle \Psi | \psi_j^\dagger\psi_j|\Psi\rangle$.
Averaging over the flux values, we obtain Eq.~(\ref{eq: current spec flow}), which is the result we set out to obtain in this subsection.


\subsection{Quantized charge pumping and the winding number}
\label{appendix: winding}

We now show that $Q_\infty$, the charged pumped over a period averaged over long times, is equal to the winding number $W$, appearing in Eq.~\eqref{eq: invariant}.  We show that in the limit of a large number of cycles, $N$, the average charge pumped per cycle contains a quantized piece equal to the winding number, plus a small correction that decays with the averaging time at least as fast as $1/N$.


Recall that the many-body initial state that we consider is a single Slater determinant with electrons populating all sites in the strip of width $\ell$ near one edge of the cylinder.
At time $t$, the expectation value of the current $\langle I_x(t)\rangle$ is given by the sum of contributions from each of these single particle states, propagated forward in time with the evolution operator $\tilde{U}\equiv \tU(\theta_x,t)$ for the system with the threaded flux $\theta_x$.
Defining a projector $\proj$ that projects onto all sites within the strip of initially occupied sites, the current is given by
\be
\label{eq: current real space no avg}
\langle I_x(t) \rangle = \textrm{Tr}\,\Big\{\tU^\dagger(\theta_x,t)\frac{\partial \tH(\theta_x,t)}{\partial\theta_x}\tU(\theta_x,t)\,\proj\Big\}.
\ee

Using Eq.~(\ref{eq: current real space no avg}), the total charge pumped in the first cycle, $\avg{Q} = \int_0^T\! dt \avg{I_x(t)}$, is given by
\begin{equation}
\avg{Q} = \int_0^T\! dt\, \textrm{Tr}\,\Big\{\tU^\dagger(\theta_x,t)\frac{\partial \tH(\theta_x,t)}{\partial\theta_x}\tU(\theta_x,t)\proj\Big\}.
\label{eq: current real space}
\end{equation}
Rearranging and using the chain rule, Eq.~(\ref{eq: current real space}) becomes
\begin{equation}
\avg{Q} = \int_0^T dt\, \textrm{Tr}\,\Big\{\proj\, \tU^{\dagger}\left(\partial_{\theta_x} (\tH\tU) - \tH \partial_{\theta_x} \tU\right)\Big\}.
\end{equation}

In the thermodynamic limit, the current $\avg{I_x(\theta_x)}$ is expected to be insensitive to the value of the threaded flux. 
Thus replacing the pumped charge by its value averaged over all $\theta_x$, and using $i\partial_t U = H U$, we obtain
\begin{equation}
\avg{Q} = \frac{i}{2\pi}\int_0^{2\pi}\! \!\!\!d\theta_x\!\int_0^T\!\! dt\, {\rm Tr} \Big\{\proj\big(\tU^{\dagger}\partial_{\theta_x}\partial_t \tU+\partial_t \tU^\dagger \partial_{\theta_x} \tU\big)\Big\},
\end{equation}
or equivalently, through integration by parts,
\begin{equation}
\avg{Q} = \frac{i}{2\pi}\int_0^{2\pi}\! \!\!\!d\theta_x\!\int_0^T\!\! dt\,\textrm{Tr}\left\{\proj\left(\partial_t U^\dagger \partial_{\theta_x} U-\partial_{\theta_x}U^{\dagger}\partial_t U\right)\right\}.
\end{equation}
Inserting $U U^\dagger = 1$ and using $(\partial_\lambda \tU^\dagger) \tU  = -\tU^\dagger \partial_\lambda \tU$ in each of the terms in the above equation gives
\begin{equation}
\avg{Q} = \frac{i}{2\pi}\int_0^{2\pi}\! \!\!\!d\theta_x\!\int_0^T\!\! dt\,\textrm{Tr}\, \Big\{(\tU^{\dagger}\partial_{\theta_x}\tU)\Big[ (\tU^{\dagger}\partial_t \tU),\proj\Big]\Big\},
\label{eq: first commutator}
\end{equation}
where 
we used $\textrm{Tr}\big\{\proj\,[A,B]\big\}=\textrm{Tr}\big\{A\,[B,\proj]\big\}$.

We now examine the cases for which the commutator in Eq.~\eqref{eq: first commutator} is non-zero.
Denoting $\tA\equiv (\tU^{\dagger}\partial_t \tU)$, the matrix element $\langle \br | [\tA,\proj] |\br'\rangle$ is nonzero in the 
 two cases
\begin{align}
\langle \br | [\tA,\proj] |\br'\rangle= -\langle \br| \tA|\br'\rangle, \quad & \proj \ket{\br} = \ket{\br}, \proj\ket{\br'} = 0\nonumber\\
\langle \br | [\tA,\proj] |\br'\rangle=\langle \br| \tA|\br'\rangle, \quad &\proj \ket{\br} = 0,\ \ \proj\ket{\br'} = \ket{\br'}. 
\label{eq: comm A Q}
\end{align}

To set up a convenient means for enforcing the conditions above, we introduce an auxillary gauge transformation under which the single particle states on the sites $\ket{\br} \equiv |x,y\rangle$ transform as
\begin{align}
&\ket{\br} \to \ket{\br}, &y& < \ell\nonumber\\ 
&\ket{\br} \to e^{i\theta_y} \ket{\br},  &\ell \leq  y& \leq L_y.  
\label{eq: gauge trans}
\end{align}
We denote the unitary operator that applies this gauge transformation as $G_{\theta_y}$.
Because Eq.~(\ref{eq: gauge trans}) defines a pure gauge transformation, the pumped charge cannot depend on the value of $\theta_y$.
Therefore we are free to average $\avg{Q}$ over all such gauges.
Using $[G_{\theta_y},\proj]=0$ and defining $\tA(\theta_y)\equiv G^\dagger_{\theta_y}(\tU^{\dagger}\partial_t \tU)G_{\theta_y}$ and $\tB(\theta_y)\equiv G^\dagger_{\theta_y}(\tU^{\dagger}\partial_{\theta_x} \tU)G_{\theta_y}$, we thus obtain 
\begin{equation}
\avg{Q} = \frac{i}{4\pi^2}\int_0^{2\pi}\! \!\!\!d\theta_y\int_0^{2\pi}\! \!\!\!d\theta_x\!\int_0^T\!\! dt\,\textrm{Tr}\Big\{\tB(\theta_y)\Big[ \tA(\theta_y),\proj\Big]\Big\}.
\end{equation}
Importantly, Eqs.~(\ref{eq: comm A Q}) and \eqref{eq: gauge trans} can be expressed as 
\begin{equation}
[\tA(\theta_y),\proj] = i \partial_{\theta_y}\tA(\theta_y),
\end{equation}
whereby the pumped charge becomes 
\begin{equation}
\avg{Q} = -\frac{1}{4\pi^2} \int_0^{2\pi}\! \!\!\!d\theta_y\int_0^{2\pi}\! \!\!\!d\theta_x\!\int_0^T\!\! dt\,\textrm{Tr}\Big\{\tB(\theta_y)\partial_{\theta_y} \tA(\theta_y)\Big\}.
\label{eq: current A B}
\end{equation}

So far, we have expressed the average current using the evolution $\tU$ on the cylinder.
Here we aim to obtain a bulk-boundary correspondence, relating the pumped charge to the evolution operator on a {\it torus}, i.e., a geometry without edges. We consider a completion of the cylinder to a torus with fluxes $\theta_x$ and $\theta_y$ threaded through the two holes of the torus. The torus Hamiltonian, $H\equiv H(\theta_x,\theta_y,t)$, is identical to  $G^\dagger_{\theta_y}\tH(\theta_x,t)G_{\theta_y}$ in the interior of the cylinder. The corresponding evolution operator is denoted by $U \equiv U(\theta_x,\theta_y,t)$.

Importantly, $U$ and $A=U^{\dagger}\partial_t U$, as well as $\tilde{U}$ and $\tilde{A}=\tilde{U}^{\dagger}\partial_t \tilde{U}$, are \textit{local} operators for $0<t<T$.
Therefore, up to corrections which are exponentially suppressed in the size of the system, $i \partial_{\theta_y}A=i \partial_{\theta_y}\tA(\theta_y)$, i.e., the derivative with respect to $\theta_y$ gives an identical result in the case of a torus and a cylinder.
Moreover, since $i \partial_{\theta_y}\tilde{A}(\theta_y)$ is also a local operator, the only matrix elements of $\langle\br'|\tB(\theta_y)|\br\rangle$ contributing in Eq.~\eqref{eq: current A B} are those for which $\br$ and $\br'$ are in the interior of the cylinder and close to the edge of the initially filled strip, i.e., $y \approx \ell$. 
For these matrix elements, $\tB$ (defined on the cylinder) and $B\equiv U^{\dagger}\partial_{\theta_x} U$ (defined on the torus) are identical (up to corrections which are exponentially small in the size of the system).
Therefore, in Eq.~\eqref{eq: current A B} we can replace $\tA$ and $\tB$ with $A$ and $B$, giving 
%
%
%
\begin{align}
\label{eq:torus}
\avg{Q} = -\frac{1}{4\pi^2}\!\oint\!  d\Theta\!\!\int_0^T\!\!\!\! dt\,\textrm{Tr}\Big\{(U^{\dagger}\partial_{\theta_x} U)\,\partial_{\theta_y}(U^{\dagger}\partial_t U)\Big\},
\end{align}
where for brevity we denoted $d\theta_xd\theta_y=d\Theta$, and united the integrals under a single integral sign.

Inserting  $U U^\dagger = 1$ between the two terms in the trace in the equation above, and again using $(\partial_\lambda U^\dagger) U  = -U^\dagger \partial_\lambda U$, we get
\begin{align}
\avg{Q} = -\frac{1}{4\pi^2}\!\oint\!  d\Theta\!\!\int_0^T\!\!\!\! dt\,\textrm{Tr}\Big\{(U^{\dagger}&\partial_{\theta_x} U)\Big(U^\dagger \partial_{\theta_y} \partial_t U \\
&-(U^\dagger\partial_{\theta_y} U)(U^\dagger\partial_{t} U) \Big)\Big\}\nonumber.
\end{align}
Using
\begin{align}
U^\dagger \partial_{\theta_y} \partial_t U &= \partial_t (U^\dagger \partial_{\theta_y} U)-\partial_t U^\dagger \partial_{\theta_y} U\nonumber\\
&= (U^\dagger \partial_t U)(U^\dagger \partial_{\theta_y} U)+\partial_t (U^\dagger \partial_{\theta_y} U),
\end{align} we get
\begin{align}
\avg{Q} = \frac{1}{4\pi^2}\!\oint\!  d\Theta\!\!\int_0^T\!\!\!\! dt\,&\textrm{Tr}\Big\{(U^{\dagger}\partial_t U)\big[(U^\dagger \partial_{\theta_x} U),\,(U^\dagger\partial_{\theta_y} U )\big]\nonumber\\
&-(U^{\dagger}\partial_{\theta_x} U)\partial_t (U^\dagger \partial_{\theta_y} U) \Big\}.
\label{eq: current almost there}
\end{align}

For the moment we focus on the second term in the above equation.
Integrating by parts gives
\begin{align}
\label{eq: last term}
&\oint\!  d\Theta\!\!\int_0^T\!\!\!\! dt\, \textrm{Tr}\Big\{(U^{\dagger}\partial_{\theta_x} U)\partial_t (U^\dagger \partial_{\theta_y} U) \Big\}\\
&=\oint\!  d\Theta\!\!\int_0^T\!\!\!\! dt\, \textrm{Tr}\Big\{\half\partial_t \Big((U^{\dagger}\partial_{\theta_x} U) (U^\dagger \partial_{\theta_y} U)\Big) \nonumber\\
&-\half(\partial_t U^{\dagger}\partial_{\theta_x} U) (U^\dagger \partial_{\theta_y} U)-\half(U^{\dagger}\partial_t \partial_{\theta_x} U) (U^\dagger \partial_{\theta_y} U)\nonumber\\
&+\half(U^{\dagger}\partial_{\theta_x} U)(\partial_t U^\dagger \partial_{\theta_y} U)+\half(U^{\dagger}\partial_{\theta_x} U)( U^\dagger \partial_t\partial_{\theta_y} U) \Big\}.\nonumber
\end{align}
Using the cyclic property of the trace, $U^\dagger U=1$, and integration by parts with respect to $\theta_x$ and $\theta_y$, it is possible to show that the third and fifth terms (containing the double derivatives) cancel. 
The second and fourth terms, using $U \partial_\lambda U^\dagger= -\partial_\lambda U U^\dagger$, can be shown to give an identical contribution to the first term in Eq.~\eqref{eq: current almost there}, but with a factor of $-\half$.
Defining the 
functional $W[U(t)]$ for a bulk evolution $U(t)$ as 
\begin{equation}
W\left[U\right]= \!\oint\!\!  \frac{d\Theta}{8\pi^2}\!\!\int_0^T\!\!\!\! dt\,\textrm{Tr}\Big\{(U^{\dagger}\partial_t U)\Big[(U^\dagger \partial_{\theta_x} U),(U^\dagger\partial_{\theta_y} U )\Big]\Big\}, 
\label{eq: winding number def}
\end{equation}
the 
net charge pumped during one driving cycle, assuming initial filling of a strip of sites covering one edge, is given by
%
%
\begin{equation}
\avg{Q} = W\left[U\right]-\! \oint\!\!  \frac{d\Theta}{8\pi^2}\!\!\int_0^T\!\!\!\! dt\,\partial_t\textrm{Tr}\Big\{(U^{\dagger}\partial_{\theta_x} U)(U^\dagger \partial_{\theta_y} U)\!\Big\}.
\label{eq: current final}
\end{equation}

It is important to note that $W[U]$ is quantized (and equal to a winding number as discussed in Ref.~\cite{Rudner2013}) only for the case where the evolution is periodic, satisfying $U(T) = U(0)$. 
For such ``ideal evolutions,'' 
the second term in Eq.~\eqref{eq: current final} clearly vanishes, and therefore the pumped charge 
is quantized and given by the winding number.

For a ``non-ideal evolution,'' where $U(T)\neq U(0)$, $W[U]$ need not be an integer.
However, if the initially-filled strip near the edge is wide enough such that all edge states are occupied with probabilities exponentially close to 1, then the spectral flow arguments presented in Sec.~\ref{appendix: spectral flow} 
indicate that the average charge pumped per cycle 
will yield a quantized value, with a correction that vanishes at least as fast as $1/N$.
As we now show, this behavior can be seen directly through further manipulations of Eq.~\eqref{eq: current final}.

Consider a ``continued'' 
evolution 
$\hat{U}(t)$, defined on a larger time period of $2T$.
We define $\hat{U}(t)$ such that it is equal to the original evolution operator $U(t)$ for $0\leq t \leq T$, and to $e^{iH_{\mathrm{eff}}(t-T)}U(T) = e^{iH_{\mathrm{eff}}(t-2T)}$ for $T<t\le2T$.
As in Ref.~\cite{Rudner2013}, $H_{\mathrm{eff}}=\frac{i}{T}\log U(T)$; in the discussion below the choice of the branch cut of the $\log$ is unimportant, as long as the system is in a localized phase.
As constructed, $\hat{U}(t)$ is an ``ideal'' evolution in the larger period $2T$, i.e., $\hat{U}(2T)=\hat{U}(0) = 1$.

Starting with Eq.~\eqref{eq: current final}, we add and subtract the quantity $W\big[e^{iH_{\mathrm{eff}}(t-T)}\big]$, i.e., Eq.~\eqref{eq: winding number def} with $U(t)$ replaced by $e^{iH_{\mathrm{eff}}(t-T)}$.
After a shift of the time variable by $T$, the added piece combines with $W[U]$ to give $W_2[\hat{U}]$, where $W_2$ is defined as in Eq.~\eqref{eq: winding number def} with the time integration taken from 0 to $2T$. 
The subtracted piece remains as a correction.
The charge pumped over one  cycle is then
\begin{align}
\avg{Q} =\ & W_2\big[\hat{U}\big]-W\big[e^{iH_{\mathrm{eff}}(t-T)}\big]\nonumber\\
&-\frac{1}{8\pi^2} \!\oint\!  d\Theta\, \textrm{Tr}\Big\{ (U^{\dagger}\partial_{\theta_x} U)(U^\dagger \partial_{\theta_y} U) \Big\}\Big|_0^T,
\label{eq: current two windings}
\end{align}
where 
the last term arises from the full derivative (second term) in Eq.~\eqref{eq: current final}.
Crucially, because $\hat{U}$ is $2T$-periodic, $W_2[\hat{U}]$ is a true winding number and is quantized.
The issue remains to characterize the contributions of the second and third terms; below we show that they 
can be neglected in the limit of a large number of pumping cycles.

Consider the average 
charge pumped over $N$ driving cycles,
$\frac{\avg{Q}_{NT}}{N} = \frac{1}{N}\int_0^{NT}\!\! dt\, \langle I_x(t)\rangle$. 
To analyze this quantity we repeat the manipulations leading up to Eq.~(\ref{eq: current final}) above.
In moving from the evolution operator on the cylinder to that on a torus, see discussion above Eq.~(\ref{eq:torus}), we furthermore use the fact that in the localized phase $U$ and $A=U\partial_t U$ remain local even at long times. Likewise, $\tilde{U}$ and $\tilde{A}=\tilde{U}\partial_t \tilde{U}$ (for the cylinder) are local in the $y$ coordinate.
In this way we find 
\begin{equation}
  \frac{\avg{Q}_{NT}}{N} = \frac{1}{N}W_N\big[U\big] + \frac{f(N)}{N},
\label{eq: current long av}
\end{equation}
where $W_N[\cdot]$ is defined in the same way as $W[\cdot]$ in Eq.~(\ref{eq: winding number def}), but with the time integration taken up to $NT$ rather than $T$.
The 
factor $f(N)$ on the right hand side of this equation 
arises from the term corresponding to the full derivative 
term in Eq.~\eqref{eq: current final}; its magnitude is bounded, and therefore the ratio $f(N)/N$ decays to zero as $N$ goes to infinity.

To see how the last term in Eq.~(\ref{eq: current long av}) vanishes for large $N$, consider $U$ in terms of its spectral decomposition, $U(NT) = \sum_n e^{-i\varepsilon_n NT}P_n$, where $P_n$ is the projector onto Floquet state $n$.
Each derivative contributes two terms: $U^\dagger \partial_{\theta_j} U = -i(\partial_{\theta_j}\varepsilon_n)  N T P_n + P_n \partial_{\theta_j} P_n$.
We consider each of the four resulting terms from the product ${\rm Tr}[(U^\dagger \partial_{\theta_x} U) (U^\dagger \partial_{\theta_y} U)]$ separately.

First, when both derivatives act on the quasienergies, we get $(NT)^2\sum_n \partial_{\theta_x}\varepsilon_n \,\partial_{\theta_y}\varepsilon_n$.
A nonzero value for these terms would imply the existence of a current that grows linearly in time, which is unphysical.
Moreover, as shown in Appendix \ref{appendix: spectral flow}, as a general rule the time-averaged current (or pumped charge) must limit to a constant plus a correction that decreases at least inversely with time.
In the fully localized phase, the quasienergies $\{\varepsilon_n\}$ are exponentially insensitive to changes in the fluxes $\theta_x$ and $\theta_y$ (see arguments below), and therefore these terms clearly give a vanishing contribution in the thermodynamic limit. 
Therefore these terms can (and must) be dropped within the level of all other approximations of exponential accuracy employed above.

Next, when one of the derivatives acts on the quasi-energy and the other acts on a projector, we get terms like $NT \sum_n \partial_{\theta_x}\varepsilon_n \, {\rm Tr}[P_n \partial_{\theta_y}P_n]$.
These terms strictly vanish due to the general identity ${\rm Tr}[P \frac{dP}{d\lambda}] = 0$, for any parameter $\lambda$ upon which the projector $P$ depends.

Finally, when both derivatives act on the projectors we get a nonvanishing contribution of the form $\sum_{n,m} {\rm Tr}[(P_n \partial_{\theta_x}P_n) (P_m \partial_{\theta_y}P_m)]$.
Crucially, these terms {\it do not depend on the length of the averaging interval, $NT$}.
Therefore the 
quantity $f(N)$ in Eq.~(\ref{eq: current long av}) is in fact constant in $N$, and the ratio $f(N)/N$ decays to zero in the long time (large $N$) limit.
Furthermore, in the localized phase, the contributions from projectors onto states localized far from the bonds where the gauge fields $\theta_{x,y}$ act are exponentially suppressed.
Thus it is clear that for any fixed $N$ the quantity $f(N)/N$ remains finite in the thermodynamic limit $L_x, L_y \rightarrow \infty$.
%

To evaluate $W_N\big[U\big]$ in Eq.~(\ref{eq: current long av}) we break up the integral over the range $0 \le t \le NT$ into $N$ segments of length $T$.
Shifting the time variable within each segment to run between 0 and $T$, we obtain 
\begin{equation}
W_N\big[U\big] = \sum_{n=0}^{N-1} W\big[U_n\big],\quad U_n(t) = U(t) U(nT).
\end{equation}
As discussed for the ``non-ideal'' evolution $U(t)$ above, the operators $\{U_n\}$ are not periodic in time and therefore $W[U_n]$ is not quantized.

To isolate the quantized contribution to Eq.~(\ref{eq: current long av}) we add and subtract a ``return map'' contribution $W[e^{iH_{\rm eff}(t-T)}U(nT)]$ for each term $W[U_n]$.
We further define the ``continued'' evolution $\hat{U}_n(t) = \hat{U}(t)U(nT)$, with $\hat{U}(t)$ as given above. 
Note that $\hat{U}_n$  is periodic in time with period $2T$ (though $\hat{U}_n(0) = \hat{U}_n(2T) \neq 1$), 
and therefore $W_2[\hat{U}_n]$ is separately quantized for each $n$.
Moreover, by virtue of the fact that the winding number $W_2[\hat{U}_n]$ is a topological invariant for periodic evolutions, its value cannot change under smooth deformations of $\hat{U}_n$.
In particular, we may deform $\hat{U}_n \rightarrow \hat{U}$ via the continuous transformation $\hat{U}_n(t; s) = \hat{U}(t)U((1-s)nT)$, by taking $s$ from 0 to 1.
Hence we find that $W_2[\hat{U}_n] = W_2[\hat{U}]$, and therefore $\sum_{n = 0}^{N-1} W_2[\hat{U}_n] = N W_2[\hat{U}]$.

Inserting the result above into Eq.~(\ref{eq: current long av}) and subtracting the appropriate return map contribution, we obtain
\begin{equation}
  \frac{\avg{Q}_{NT}}{N} = W_2[\hat{U}] - \frac{1}{N}W_N\big[e^{iH_{\mathrm{eff}}(t-NT)}\big] + \frac{f(N)}{N},
\label{eq: current long av2}
\end{equation}
where we have combined the contributions of the return maps for all $n$ into one term $W_N\big[e^{iH_{\mathrm{eff}}(t-NT)}\big]$.
Note that by shifting time arguments we can make the replacement $W_N\big[e^{iH_{\mathrm{eff}}(t-NT)}\big] = -W_N\big[e^{-iH_{\mathrm{eff}}t}\big]$.


The quantity $W_N[e^{-iH_{\mathrm{eff}}t}]$ is not necessarily quantized, 
since the unitary $e^{-iH_{\mathrm{eff}}t}$ is not a periodic function of $t$ over the range $0 \le t \le NT$.
However, if the eigenstates of $H_{\mathrm{eff}}$ are all localized, 
then we can show (see below) that   $W_N[e^{-iH_{\mathrm{eff}}t}]$ decays with $N$ as $~1/N$ (or faster).
The underlying reason is that, in the localized case, 
the eigenstates of $H_{\mathrm{eff}}$ do not flow under insertion of the fluxes $\theta_x$ and $\theta_y$ into the torus.
For now we simply assert this claim, and will prove it at the end of this section. 
Accepting the claim to be true, we obtain
\begin{equation}
\frac{\avg{Q}_{NT}}{N} = W_2[\hat{U}]+\tilde{f}(N)/N,
\label{eq: current final result}
\end{equation}
where $\tilde{f}(N)$ is bounded by a constant as a function of $N$. Equation~\eqref{eq: current final result} is the result we set out to prove in this Appendix.

Finally, to close the loose ends, we show that $W_N\left[e^{-iH_{\mathrm{eff}}t}\right]$ is  bounded as a function of $N$.
Using the spectral decomposition $e^{-iH_{\mathrm{eff}}t}=\sum_n e^{-i\varepsilon_n t} P_n$, we have
\begin{align}
&W_N\left[e^{-iH_{\mathrm{eff}}t}\right]\nonumber\\&=\sum \!\!\!\int_0^{NT}\!\!\!\!\!\!\!\! d\Theta dt\, \frac{e^{-i\Delta\varepsilon t}\varepsilon_n}{i8\pi^2}\,\textrm{Tr}\Big\{P_n[P_{m_1}\partial_{\theta_x} P_{m_2},P_{k_1}\partial_{\theta_y} P_{k_2}]\Big\},
\label{eq: w heff spectral}
\end{align}
with $\Delta\varepsilon=\varepsilon_{m_1}+\varepsilon_{k_1}-\varepsilon_{m_2}-\varepsilon_{k_2}$, and the sum taken over the integers $n,m_{1,2},k_{1,2}$. To get to Eq.~\eqref{eq: w heff spectral}, we used $\textrm{Tr}\Big\{P_n[P_{m},P_{k}]\Big\}=0$ and $\textrm{Tr}\Big\{P_n[P_{m},P_{k_1}\partial_{\theta_y} P_{k_2}]\Big\}=0$.

When $H_{\mathrm{eff}}$ is fully localized, its eigenstates do not ``wrap'' around the cycles of the torus, and therefore are  insusceptible to the flux insertion. Therefore, up to corrections which are suppressed as $\exp(-L/\xi)$, where $\xi$ is the localization length, we have that:  (i) the eigenvalues $\varepsilon_n(\Theta)$ are independent of the values of the fluxes; (ii) under changing the values for the fluxes,  the projectors $P_n$ transform as if transforming under a local gauge transformation, $P_n(\Theta)=e^{i\Upsilon}P_n e^{-i\Upsilon}$, with $e^{i\Upsilon}=e^{i\mathcal{Q}_x\theta_x+\mathcal{Q}_y\theta_y}$. Here, $\mathcal{Q}_x$ and $\mathcal{Q}_y$ are projectors on sites that define the gauge transformation felt by the localized eigenstates.

Returning to Eq.~\eqref{eq: w heff spectral}, we note  that in order for a term in Eq.~\eqref{eq: w heff spectral} to 
grow with $N$, it must have $\Delta\varepsilon=0$. Excluding the possibility of a fine tuned degeneracy which occurs on a finite area in flux space, the condition $\Delta\varepsilon=0$  requires that either $m_1=m_2, k_1=k_2$ or $m_1=k_2, m_1=k_1$. However, the contribution of the latter two cases to $W_N\left[e^{-iH_{\mathrm{eff}}t}\right]$ can be shown to vanish by substituting $\partial_{\theta_\alpha} P_m= ie^{i\Upsilon}\left[\mathcal{Q}_\alpha,P_m\right]e^{-i\Upsilon}$ in Eq.~\eqref{eq: w heff spectral}, and applying straightforward algebraic manipulations. Therefore, we find that all the terms in $W_N\left[e^{-iH_{\mathrm{eff}}t}\right]$ are bounded by a constant as a function of $N$, and thereby we obtain Eq.~\eqref{eq: current long av2}.

\section{Charge pumping statistics}
\label{appendix: single pump}
\begin{figure}
\includegraphics[width=1.0\linewidth]{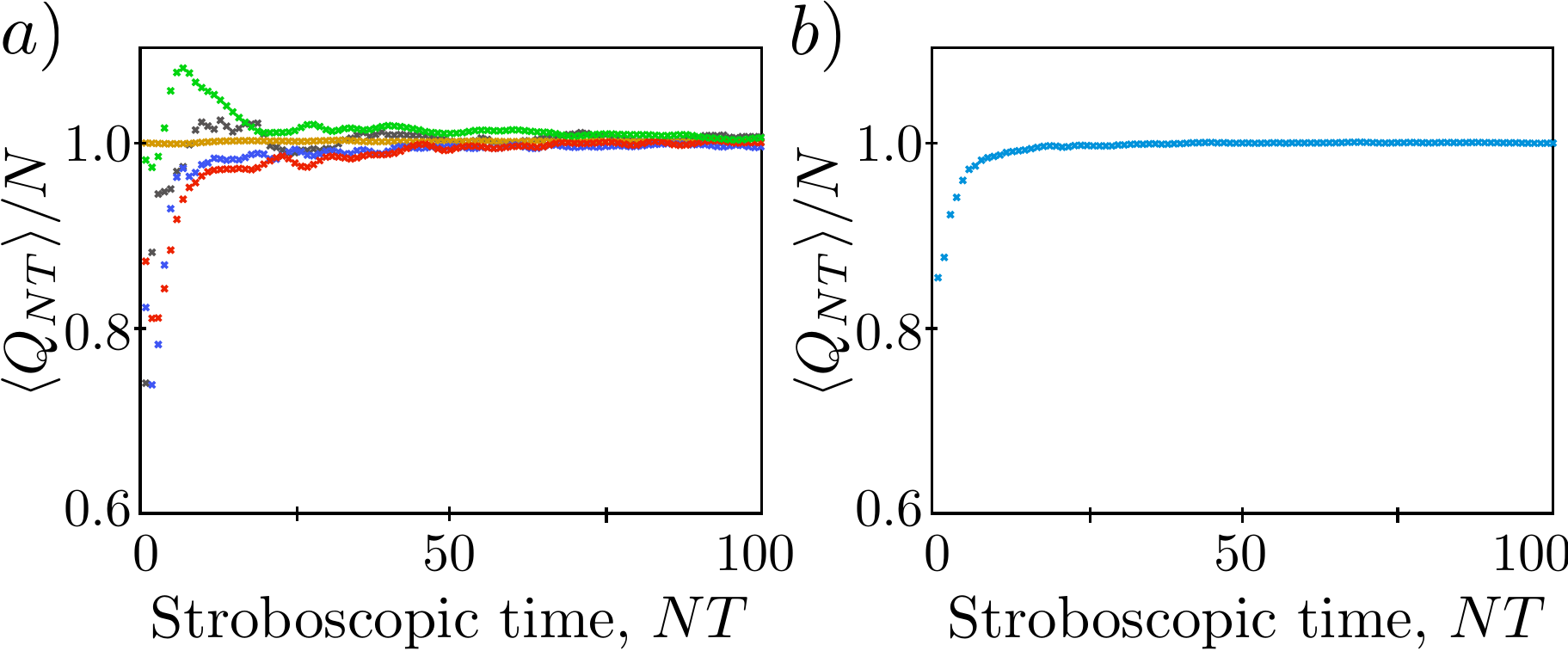}
\caption{Quantized charge pumping in the AFAI in a single disorder realization for a system of size $50\times50$. All parameters are the same as in Fig. \ref{fig:chargepumping}. (a) Cumulative average of pumped charge per cycle, $Q_{\infty}$, resolved for five different vertical cuts across the cylinder (cuts along the $y$ direction, c.f. Fig.~\ref{fig: cylinder}). The spread of asymptotic values arises from numerical error in the finite time differences used for evaluating the integral in Eq.~\eqref{eq: current def}. (b) The pumped  charge, averaged over all longitudinal cuts in the same finite-sized system. For a thermodynamically large system, averaging over all cuts is equivalent to averaging over disorder realizations, see Fig.~\ref{fig:chargepumping}(b).}
\label{fig:pumpingstats}
\end{figure}

As mentioned in the main text, quantization of the charge pumped per cycle is realized for every individual disorder realization in a large system.  In this short appendix we show numerical results for a \textit{single} disorder realization in a finite system of size $50 \times 50$ lattice sites.  In the main text we defined the pumped charge by integrating the current across a single vertical``cut" across the system in a cylinder geometry (a cut along the $y$ direction, c.f. Fig.~\ref{fig: cylinder}).  Here, in Fig.~\ref{fig:pumpingstats} we show that the detailed time dependence of $\avg{Q}_{NT}$, the cumulative average charge pumped per cycle across each \textit{single} cut, displays a unique pattern of decaying oscillations and limits to one at large times (panel a).  Current conservation implies that the average currents across all cuts must be equal in the long time limit; the spread of values at large times is due to numerical discretization error.  In Fig.~\ref{fig:pumpingstats}(b) we show the current averaged over all vertical cuts in the same $50\times50$ system.  Here the rapid convergence to the quantized value is clearly displayed.

\bibliographystyle{apsrev}
\bibliography{zero_chern_refs}

\begin{thebibliography}{53}
\expandafter\ifx\csname natexlab\endcsname\relax\def\natexlab#1{#1}\fi
\expandafter\ifx\csname bibnamefont\endcsname\relax
  \def\bibnamefont#1{#1}\fi
\expandafter\ifx\csname bibfnamefont\endcsname\relax
  \def\bibfnamefont#1{#1}\fi
\expandafter\ifx\csname citenamefont\endcsname\relax
  \def\citenamefont#1{#1}\fi
\expandafter\ifx\csname url\endcsname\relax
  \def\url#1{\texttt{#1}}\fi
\expandafter\ifx\csname urlprefix\endcsname\relax\def\urlprefix{URL }\fi
\providecommand{\bibinfo}[2]{#2}
\providecommand{\eprint}[2][]{\url{#2}}

\bibitem[{\citenamefont{Oka and Aoki}(2009)}]{Oka2009}
\bibinfo{author}{\bibfnamefont{T.}~\bibnamefont{Oka}} \bibnamefont{and}
  \bibinfo{author}{\bibfnamefont{H.}~\bibnamefont{Aoki}},
  \bibinfo{journal}{Phys. Rev. B} \textbf{\bibinfo{volume}{79}},
  \bibinfo{pages}{081406} (\bibinfo{year}{2009}).

\bibitem[{\citenamefont{Inoue and Tanaka}(2010)}]{Inoue2010}
\bibinfo{author}{\bibfnamefont{J.-i.} \bibnamefont{Inoue}} \bibnamefont{and}
  \bibinfo{author}{\bibfnamefont{A.}~\bibnamefont{Tanaka}},
  \bibinfo{journal}{Phys. Rev. Lett.} \textbf{\bibinfo{volume}{105}},
  \bibinfo{pages}{017401} (\bibinfo{year}{2010}).

\bibitem[{\citenamefont{Kitagawa et~al.}(2010)\citenamefont{Kitagawa, Berg,
  Rudner, and Demler}}]{KBRD}
\bibinfo{author}{\bibfnamefont{T.}~\bibnamefont{Kitagawa}},
  \bibinfo{author}{\bibfnamefont{E.}~\bibnamefont{Berg}},
  \bibinfo{author}{\bibfnamefont{M.}~\bibnamefont{Rudner}}, \bibnamefont{and}
  \bibinfo{author}{\bibfnamefont{E.}~\bibnamefont{Demler}},
  \bibinfo{journal}{Phys. Rev. B} \textbf{\bibinfo{volume}{82}},
  \bibinfo{pages}{235114} (\bibinfo{year}{2010}).

\bibitem[{\citenamefont{Lindner et~al.}(2011)\citenamefont{Lindner, Refael, and
  Galitski}}]{Lindner2011}
\bibinfo{author}{\bibfnamefont{N.~H.} \bibnamefont{Lindner}},
  \bibinfo{author}{\bibfnamefont{G.}~\bibnamefont{Refael}}, \bibnamefont{and}
  \bibinfo{author}{\bibfnamefont{V.}~\bibnamefont{Galitski}},
  \bibinfo{journal}{Nat. Phys.} \textbf{\bibinfo{volume}{7}},
  \bibinfo{pages}{490} (\bibinfo{year}{2011}).

\bibitem[{\citenamefont{Lindner et~al.}(2013)\citenamefont{Lindner, Bergman,
  Refael, and Galitski}}]{Lindner2013}
\bibinfo{author}{\bibfnamefont{N.~H.} \bibnamefont{Lindner}},
  \bibinfo{author}{\bibfnamefont{D.~L.} \bibnamefont{Bergman}},
  \bibinfo{author}{\bibfnamefont{G.}~\bibnamefont{Refael}}, \bibnamefont{and}
  \bibinfo{author}{\bibfnamefont{V.}~\bibnamefont{Galitski}},
  \bibinfo{journal}{Phys. Rev. B} \textbf{\bibinfo{volume}{87}},
  \bibinfo{pages}{235131} (\bibinfo{year}{2013}).

\bibitem[{\citenamefont{Gu et~al.}(2011)\citenamefont{Gu, Fertig, Arovas, and
  Auerbach}}]{Gu11}
\bibinfo{author}{\bibfnamefont{Z.}~\bibnamefont{Gu}},
  \bibinfo{author}{\bibfnamefont{H.~A.} \bibnamefont{Fertig}},
  \bibinfo{author}{\bibfnamefont{D.~P.} \bibnamefont{Arovas}},
  \bibnamefont{and} \bibinfo{author}{\bibfnamefont{A.}~\bibnamefont{Auerbach}},
  \bibinfo{journal}{Phys. Rev. Lett.} \textbf{\bibinfo{volume}{107}},
  \bibinfo{pages}{216601} (\bibinfo{year}{2011}).

\bibitem[{\citenamefont{Kitagawa et~al.}(2011)\citenamefont{Kitagawa, Oka,
  Brataas, Fu, and Demler}}]{Kitagawa2011}
\bibinfo{author}{\bibfnamefont{T.}~\bibnamefont{Kitagawa}},
  \bibinfo{author}{\bibfnamefont{T.}~\bibnamefont{Oka}},
  \bibinfo{author}{\bibfnamefont{A.}~\bibnamefont{Brataas}},
  \bibinfo{author}{\bibfnamefont{L.}~\bibnamefont{Fu}}, \bibnamefont{and}
  \bibinfo{author}{\bibfnamefont{E.}~\bibnamefont{Demler}},
  \bibinfo{journal}{Phys. Rev. B} \textbf{\bibinfo{volume}{84}},
  \bibinfo{pages}{235108} (\bibinfo{year}{2011}).

\bibitem[{\citenamefont{Delplace et~al.}(2013)\citenamefont{Delplace,
  G\'omez-Le\'on, and Platero}}]{Delplace2013}
\bibinfo{author}{\bibfnamefont{P.}~\bibnamefont{Delplace}},
  \bibinfo{author}{\bibfnamefont{A.}~\bibnamefont{G\'omez-Le\'on}},
  \bibnamefont{and} \bibinfo{author}{\bibfnamefont{G.}~\bibnamefont{Platero}},
  \bibinfo{journal}{Phys. Rev. B} \textbf{\bibinfo{volume}{88}},
  \bibinfo{pages}{245422} (\bibinfo{year}{2013}).

\bibitem[{\citenamefont{Katan and Podolsky}(2013)}]{Podolsky2013}
\bibinfo{author}{\bibfnamefont{Y.~T.} \bibnamefont{Katan}} \bibnamefont{and}
  \bibinfo{author}{\bibfnamefont{D.}~\bibnamefont{Podolsky}},
  \bibinfo{journal}{Phys. Rev. Lett.} \textbf{\bibinfo{volume}{110}},
  \bibinfo{pages}{016802} (\bibinfo{year}{2013}).

\bibitem[{\citenamefont{Titum et~al.}(2015)\citenamefont{Titum, Lindner,
  Rechtsman, and Refael}}]{Titum2015}
\bibinfo{author}{\bibfnamefont{P.}~\bibnamefont{Titum}},
  \bibinfo{author}{\bibfnamefont{N.~H.} \bibnamefont{Lindner}},
  \bibinfo{author}{\bibfnamefont{M.~C.} \bibnamefont{Rechtsman}},
  \bibnamefont{and} \bibinfo{author}{\bibfnamefont{G.}~\bibnamefont{Refael}},
  \bibinfo{journal}{Phys. Rev. Lett.} \textbf{\bibinfo{volume}{114}},
  \bibinfo{pages}{056801} (\bibinfo{year}{2015}).

\bibitem[{\citenamefont{Usaj et~al.}(2014)\citenamefont{Usaj, Perez-Piskunow,
  Foa~Torres, and Balseiro}}]{TorresPRB2014}
\bibinfo{author}{\bibfnamefont{G.}~\bibnamefont{Usaj}},
  \bibinfo{author}{\bibfnamefont{P.~M.} \bibnamefont{Perez-Piskunow}},
  \bibinfo{author}{\bibfnamefont{L.~E.~F.} \bibnamefont{Foa~Torres}},
  \bibnamefont{and} \bibinfo{author}{\bibfnamefont{C.~A.}
  \bibnamefont{Balseiro}}, \bibinfo{journal}{Phys. Rev. B}
  \textbf{\bibinfo{volume}{90}}, \bibinfo{pages}{115423}
  (\bibinfo{year}{2014}).

\bibitem[{\citenamefont{Foa~Torres et~al.}(2014)\citenamefont{Foa~Torres,
  Perez-Piskunow, Balseiro, and Usaj}}]{TorresPRL2014}
\bibinfo{author}{\bibfnamefont{L.~E.~F.} \bibnamefont{Foa~Torres}},
  \bibinfo{author}{\bibfnamefont{P.~M.} \bibnamefont{Perez-Piskunow}},
  \bibinfo{author}{\bibfnamefont{C.~A.} \bibnamefont{Balseiro}},
  \bibnamefont{and} \bibinfo{author}{\bibfnamefont{G.}~\bibnamefont{Usaj}},
  \bibinfo{journal}{Phys. Rev. Lett.} \textbf{\bibinfo{volume}{113}},
  \bibinfo{pages}{266801} (\bibinfo{year}{2014}).

\bibitem[{\citenamefont{{D'Alessio} and {Rigol}}(2014)}]{AlessioArxiv2014}
\bibinfo{author}{\bibfnamefont{L.}~\bibnamefont{{D'Alessio}}} \bibnamefont{and}
  \bibinfo{author}{\bibfnamefont{M.}~\bibnamefont{{Rigol}}},
  \bibinfo{journal}{arXiv:1409.6319}  (\bibinfo{year}{2014}).

\bibitem[{\citenamefont{Dehghani
  et~al.}(2014{\natexlab{a}})\citenamefont{Dehghani, Oka, and
  Mitra}}]{Dehghani2014}
\bibinfo{author}{\bibfnamefont{H.}~\bibnamefont{Dehghani}},
  \bibinfo{author}{\bibfnamefont{T.}~\bibnamefont{Oka}}, \bibnamefont{and}
  \bibinfo{author}{\bibfnamefont{A.}~\bibnamefont{Mitra}},
  \bibinfo{journal}{Phys. Rev. B} \textbf{\bibinfo{volume}{90}},
  \bibinfo{pages}{195429} (\bibinfo{year}{2014}{\natexlab{a}}).

\bibitem[{\citenamefont{Dehghani
  et~al.}(2014{\natexlab{b}})\citenamefont{Dehghani, Oka, and
  Mitra}}]{Dehghani2014b}
\bibinfo{author}{\bibfnamefont{H.}~\bibnamefont{Dehghani}},
  \bibinfo{author}{\bibfnamefont{T.}~\bibnamefont{Oka}}, \bibnamefont{and}
  \bibinfo{author}{\bibfnamefont{A.}~\bibnamefont{Mitra}},
  \bibinfo{journal}{arXiv:1412.8469}  (\bibinfo{year}{2014}{\natexlab{b}}).

\bibitem[{\citenamefont{Bilitewski and Cooper}(2014)}]{Bilitewski2014}
\bibinfo{author}{\bibfnamefont{T.}~\bibnamefont{Bilitewski}} \bibnamefont{and}
  \bibinfo{author}{\bibfnamefont{N.~R.} \bibnamefont{Cooper}},
  \bibinfo{journal}{arXiv:1410.5364}  (\bibinfo{year}{2014}).

\bibitem[{\citenamefont{Sentef et~al.}(2015)\citenamefont{Sentef, Claassen,
  Kemper, Moritz, Oka, and Freericks}}]{Sentef2015}
\bibinfo{author}{\bibfnamefont{M.}~\bibnamefont{Sentef}},
  \bibinfo{author}{\bibfnamefont{M.}~\bibnamefont{Claassen}},
  \bibinfo{author}{\bibfnamefont{A.}~\bibnamefont{Kemper}},
  \bibinfo{author}{\bibfnamefont{B.}~\bibnamefont{Moritz}},
  \bibinfo{author}{\bibfnamefont{T.}~\bibnamefont{Oka}}, \bibnamefont{and}
  \bibinfo{author}{\bibfnamefont{T.}~\bibnamefont{Freericks},
  \bibfnamefont{J.Kand~Devereaux}}, \bibinfo{journal}{Nat. Comm.}
  \textbf{\bibinfo{volume}{6}}, \bibinfo{pages}{7047} (\bibinfo{year}{2015}).

\bibitem[{\citenamefont{Seetharam et~al.}(2015)\citenamefont{Seetharam, Bardyn,
  Lindner, Rudner, and Refael}}]{Seetharam2015}
\bibinfo{author}{\bibfnamefont{K.~I.} \bibnamefont{Seetharam}},
  \bibinfo{author}{\bibfnamefont{C.-E.} \bibnamefont{Bardyn}},
  \bibinfo{author}{\bibfnamefont{N.~H.} \bibnamefont{Lindner}},
  \bibinfo{author}{\bibfnamefont{M.~S.} \bibnamefont{Rudner}},
  \bibnamefont{and} \bibinfo{author}{\bibfnamefont{G.}~\bibnamefont{Refael}},
  \bibinfo{journal}{arXiv:1502.02664v1}  (\bibinfo{year}{2015}).

\bibitem[{\citenamefont{{Iadecola} et~al.}(2015)\citenamefont{{Iadecola},
  {Neupert}, and {Chamon}}}]{Iadecola2015}
\bibinfo{author}{\bibfnamefont{T.}~\bibnamefont{{Iadecola}}},
  \bibinfo{author}{\bibfnamefont{T.}~\bibnamefont{{Neupert}}},
  \bibnamefont{and} \bibinfo{author}{\bibfnamefont{C.}~\bibnamefont{{Chamon}}},
  \bibinfo{journal}{arXiv:1502.05047}  (\bibinfo{year}{2015}).

\bibitem[{\citenamefont{Wang et~al.}(2013)\citenamefont{Wang, Steinberg,
  Jarillo-Herrero, and Gedik}}]{Wang2013}
\bibinfo{author}{\bibfnamefont{Y.~H.} \bibnamefont{Wang}},
  \bibinfo{author}{\bibfnamefont{H.}~\bibnamefont{Steinberg}},
  \bibinfo{author}{\bibfnamefont{P.}~\bibnamefont{Jarillo-Herrero}},
  \bibnamefont{and} \bibinfo{author}{\bibfnamefont{N.}~\bibnamefont{Gedik}},
  \bibinfo{journal}{Science} \textbf{\bibinfo{volume}{342}},
  \bibinfo{pages}{453} (\bibinfo{year}{2013}).

\bibitem[{\citenamefont{Jotzu et~al.}(2014)\citenamefont{Jotzu, Messer,
  Desbuquois, Lebrat, Uehlinger, Greif, and Esslinger}}]{Jotzu2014}
\bibinfo{author}{\bibfnamefont{G.}~\bibnamefont{Jotzu}},
  \bibinfo{author}{\bibfnamefont{M.}~\bibnamefont{Messer}},
  \bibinfo{author}{\bibfnamefont{R.}~\bibnamefont{Desbuquois}},
  \bibinfo{author}{\bibfnamefont{M.}~\bibnamefont{Lebrat}},
  \bibinfo{author}{\bibfnamefont{T.}~\bibnamefont{Uehlinger}},
  \bibinfo{author}{\bibfnamefont{D.}~\bibnamefont{Greif}}, \bibnamefont{and}
  \bibinfo{author}{\bibfnamefont{T.}~\bibnamefont{Esslinger}},
  \bibinfo{journal}{Nature} \textbf{\bibinfo{volume}{515}},
  \bibinfo{pages}{237} (\bibinfo{year}{2014}).

\bibitem[{\citenamefont{Rechtsman et~al.}(2013)\citenamefont{Rechtsman, Zeuner,
  Plotnik, Lumer, Podolsky, Dreisow, Nolte, Segev, and
  Szameit}}]{Rechtsman2013}
\bibinfo{author}{\bibfnamefont{M.~C.} \bibnamefont{Rechtsman}},
  \bibinfo{author}{\bibfnamefont{J.~M.} \bibnamefont{Zeuner}},
  \bibinfo{author}{\bibfnamefont{Y.}~\bibnamefont{Plotnik}},
  \bibinfo{author}{\bibfnamefont{Y.}~\bibnamefont{Lumer}},
  \bibinfo{author}{\bibfnamefont{D.}~\bibnamefont{Podolsky}},
  \bibinfo{author}{\bibfnamefont{F.}~\bibnamefont{Dreisow}},
  \bibinfo{author}{\bibfnamefont{S.}~\bibnamefont{Nolte}},
  \bibinfo{author}{\bibfnamefont{M.}~\bibnamefont{Segev}}, \bibnamefont{and}
  \bibinfo{author}{\bibfnamefont{A.}~\bibnamefont{Szameit}},
  \bibinfo{journal}{Nature} \textbf{\bibinfo{volume}{496}},
  \bibinfo{pages}{196} (\bibinfo{year}{2013}).

\bibitem[{\citenamefont{Jiang et~al.}(2011)\citenamefont{Jiang, Kitagawa,
  Alicea, Akhmerov, Pekker, Refael, Cirac, Demler, Lukin, and
  Zoller}}]{Jiang2011}
\bibinfo{author}{\bibfnamefont{L.}~\bibnamefont{Jiang}},
  \bibinfo{author}{\bibfnamefont{T.}~\bibnamefont{Kitagawa}},
  \bibinfo{author}{\bibfnamefont{J.}~\bibnamefont{Alicea}},
  \bibinfo{author}{\bibfnamefont{A.~R.} \bibnamefont{Akhmerov}},
  \bibinfo{author}{\bibfnamefont{D.}~\bibnamefont{Pekker}},
  \bibinfo{author}{\bibfnamefont{G.}~\bibnamefont{Refael}},
  \bibinfo{author}{\bibfnamefont{J.~I.} \bibnamefont{Cirac}},
  \bibinfo{author}{\bibfnamefont{E.}~\bibnamefont{Demler}},
  \bibinfo{author}{\bibfnamefont{M.~D.} \bibnamefont{Lukin}}, \bibnamefont{and}
  \bibinfo{author}{\bibfnamefont{P.}~\bibnamefont{Zoller}},
  \bibinfo{journal}{Phys. Rev. Lett.} \textbf{\bibinfo{volume}{106}},
  \bibinfo{pages}{220402} (\bibinfo{year}{2011}).

\bibitem[{\citenamefont{Rudner et~al.}(2013)\citenamefont{Rudner, Lindner,
  Berg, and Levin}}]{Rudner2013}
\bibinfo{author}{\bibfnamefont{M.~S.} \bibnamefont{Rudner}},
  \bibinfo{author}{\bibfnamefont{N.~H.} \bibnamefont{Lindner}},
  \bibinfo{author}{\bibfnamefont{E.}~\bibnamefont{Berg}}, \bibnamefont{and}
  \bibinfo{author}{\bibfnamefont{M.}~\bibnamefont{Levin}},
  \bibinfo{journal}{Phys. Rev. X} \textbf{\bibinfo{volume}{3}},
  \bibinfo{pages}{031005} (\bibinfo{year}{2013}).

\bibitem[{\citenamefont{Kundu and Seradjeh}(2013)}]{Kundu2013}
\bibinfo{author}{\bibfnamefont{A.}~\bibnamefont{Kundu}} \bibnamefont{and}
  \bibinfo{author}{\bibfnamefont{B.}~\bibnamefont{Seradjeh}},
  \bibinfo{journal}{Phys. Rev. Lett.} \textbf{\bibinfo{volume}{111}},
  \bibinfo{pages}{136402} (\bibinfo{year}{2013}).

\bibitem[{\citenamefont{Carpentier et~al.}(2015)\citenamefont{Carpentier,
  Delplace, Fruchart, and Gawedzki}}]{Carpentier}
\bibinfo{author}{\bibfnamefont{D.}~\bibnamefont{Carpentier}},
  \bibinfo{author}{\bibfnamefont{P.}~\bibnamefont{Delplace}},
  \bibinfo{author}{\bibfnamefont{M.}~\bibnamefont{Fruchart}}, \bibnamefont{and}
  \bibinfo{author}{\bibfnamefont{K.}~\bibnamefont{Gawedzki}},
  \bibinfo{journal}{Phys. Rev. Lett.} \textbf{\bibinfo{volume}{114}},
  \bibinfo{pages}{106806} (\bibinfo{year}{2015}).

\bibitem[{\citenamefont{Asboth et~al.}(2014)\citenamefont{Asboth, Tarasinski,
  and Delplace}}]{Asboth2014}
\bibinfo{author}{\bibfnamefont{J.~K.} \bibnamefont{Asboth}},
  \bibinfo{author}{\bibfnamefont{B.}~\bibnamefont{Tarasinski}},
  \bibnamefont{and} \bibinfo{author}{\bibfnamefont{P.}~\bibnamefont{Delplace}},
  \bibinfo{journal}{Phys. Rev. B} \textbf{\bibinfo{volume}{90}},
  \bibinfo{pages}{125143} (\bibinfo{year}{2014}).

\bibitem[{\citenamefont{Thouless et~al.}(1982)\citenamefont{Thouless, Kohmoto,
  Nightingale, and den Nijs}}]{Thouless1982}
\bibinfo{author}{\bibfnamefont{D.~J.} \bibnamefont{Thouless}},
  \bibinfo{author}{\bibfnamefont{M.}~\bibnamefont{Kohmoto}},
  \bibinfo{author}{\bibfnamefont{M.~P.} \bibnamefont{Nightingale}},
  \bibnamefont{and} \bibinfo{author}{\bibfnamefont{M.}~\bibnamefont{den Nijs}},
  \bibinfo{journal}{Phys. Rev. Lett.} \textbf{\bibinfo{volume}{49}},
  \bibinfo{pages}{405} (\bibinfo{year}{1982}).

\bibitem[{\citenamefont{Hu et~al.}(2015)\citenamefont{Hu, Pillay, Wu, Pasek,
  Shum, and Chong}}]{Hu2015}
\bibinfo{author}{\bibfnamefont{W.}~\bibnamefont{Hu}},
  \bibinfo{author}{\bibfnamefont{J.~C.} \bibnamefont{Pillay}},
  \bibinfo{author}{\bibfnamefont{K.}~\bibnamefont{Wu}},
  \bibinfo{author}{\bibfnamefont{M.}~\bibnamefont{Pasek}},
  \bibinfo{author}{\bibfnamefont{P.~P.} \bibnamefont{Shum}}, \bibnamefont{and}
  \bibinfo{author}{\bibfnamefont{Y.~D.} \bibnamefont{Chong}},
  \bibinfo{journal}{Phys. Rev. X} \textbf{\bibinfo{volume}{5}},
  \bibinfo{pages}{011012} (\bibinfo{year}{2015}).

\bibitem[{\citenamefont{Thouless}(1984)}]{Thouless1984}
\bibinfo{author}{\bibfnamefont{D.~J.} \bibnamefont{Thouless}},
  \bibinfo{journal}{Journal of Physics C: Solid State Physics}
  \textbf{\bibinfo{volume}{17}}, \bibinfo{pages}{L325} (\bibinfo{year}{1984}).

\bibitem[{\citenamefont{Thonhauser and Vanderbilt}(2006)}]{Thonhauser2006}
\bibinfo{author}{\bibfnamefont{T.}~\bibnamefont{Thonhauser}} \bibnamefont{and}
  \bibinfo{author}{\bibfnamefont{D.}~\bibnamefont{Vanderbilt}},
  \bibinfo{journal}{Phys. Rev. B} \textbf{\bibinfo{volume}{74}},
  \bibinfo{pages}{235111} (\bibinfo{year}{2006}).

\bibitem[{\citenamefont{Halperin}(1982)}]{Halperin1982}
\bibinfo{author}{\bibfnamefont{B.~I.} \bibnamefont{Halperin}},
  \bibinfo{journal}{Phys. Rev. B} \textbf{\bibinfo{volume}{25}},
  \bibinfo{pages}{2185} (\bibinfo{year}{1982}).

\bibitem[{\citenamefont{Thouless}(1983)}]{Thouless1983}
\bibinfo{author}{\bibfnamefont{D.~J.} \bibnamefont{Thouless}},
  \bibinfo{journal}{Phys. Rev. B} \textbf{\bibinfo{volume}{27}},
  \bibinfo{pages}{6083} (\bibinfo{year}{1983}).

\bibitem[{foo()}]{footnote:gauge}
\bibinfo{note}{The winding number introduced below is independent of the gauge
  used to represent these fluxes}.

\bibitem[{not()}]{note-Chern}
\bibinfo{note}{This can be seen, e.g., from the fact that for any given
  localized state, one can choose a gauge for which this state is essentially
  independent of $\theta_x$ and $\theta_y$.}

\bibitem[{\citenamefont{Avron and Seiler}(1985)}]{Avron1985}
\bibinfo{author}{\bibfnamefont{J.~E.} \bibnamefont{Avron}} \bibnamefont{and}
  \bibinfo{author}{\bibfnamefont{R.}~\bibnamefont{Seiler}},
  \bibinfo{journal}{Phys. Rev. Lett.} \textbf{\bibinfo{volume}{54}},
  \bibinfo{pages}{259} (\bibinfo{year}{1985}).

\bibitem[{Spi()}]{Spiros2015-comment}
\bibinfo{note}{For the Hall conductance, a more careful treatments show that
  averaging over $\theta_x$ is not necessary. See: M. Spyridon and M. Hastings,
  Commun. in Math. Phys. \textbf{334}, 433 (2015).}

\bibitem[{\citenamefont{Anderson}(1958)}]{Anderson1958}
\bibinfo{author}{\bibfnamefont{P.~W.} \bibnamefont{Anderson}},
  \bibinfo{journal}{Phys. Rev.} \textbf{\bibinfo{volume}{109}},
  \bibinfo{pages}{1492} (\bibinfo{year}{1958}).

\bibitem[{\citenamefont{Mehta}(2004)}]{Mehta2004}
\bibinfo{author}{\bibfnamefont{M.}~\bibnamefont{Mehta}},
  \emph{\bibinfo{title}{Random Matrices, Pure and Applied Mathematics}}, vol.
  \bibinfo{volume}{142} (\bibinfo{publisher}{Elsevier/Academic Press},
  \bibinfo{address}{Amsterdam, Netherlands}, \bibinfo{year}{2004}),
  \bibinfo{edition}{3rd} ed.

\bibitem[{\citenamefont{Oganesyan and Huse}(2007)}]{OganesyanHuse}
\bibinfo{author}{\bibfnamefont{V.}~\bibnamefont{Oganesyan}} \bibnamefont{and}
  \bibinfo{author}{\bibfnamefont{D.~A.} \bibnamefont{Huse}},
  \bibinfo{journal}{Phys. Rev. B} \textbf{\bibinfo{volume}{75}},
  \bibinfo{pages}{155111} (\bibinfo{year}{2007}).

\bibitem[{\citenamefont{Atas et~al.}(2013)\citenamefont{Atas, Bogomolny,
  Giraud, and Roux}}]{Atas2013}
\bibinfo{author}{\bibfnamefont{Y.~Y.} \bibnamefont{Atas}},
  \bibinfo{author}{\bibfnamefont{E.}~\bibnamefont{Bogomolny}},
  \bibinfo{author}{\bibfnamefont{O.}~\bibnamefont{Giraud}}, \bibnamefont{and}
  \bibinfo{author}{\bibfnamefont{G.}~\bibnamefont{Roux}},
  \bibinfo{journal}{Phys. Rev. Lett.} \textbf{\bibinfo{volume}{110}},
  \bibinfo{pages}{084101} (\bibinfo{year}{2013}).

\bibitem[{\citenamefont{DÕAlessio and Rigol}(2014)}]{DAlessio2014}
\bibinfo{author}{\bibfnamefont{L.}~\bibnamefont{DÕAlessio}} \bibnamefont{and}
  \bibinfo{author}{\bibfnamefont{M.}~\bibnamefont{Rigol}},
  \bibinfo{journal}{Physical Review X} \textbf{\bibinfo{volume}{4}},
  \bibinfo{pages}{041048} (\bibinfo{year}{2014}).

\bibitem[{\citenamefont{Evers and Mirlin}(2008)}]{Mirlin2008}
\bibinfo{author}{\bibfnamefont{F.}~\bibnamefont{Evers}} \bibnamefont{and}
  \bibinfo{author}{\bibfnamefont{A.~D.} \bibnamefont{Mirlin}},
  \bibinfo{journal}{Rev. Mod. Phys.} \textbf{\bibinfo{volume}{80}},
  \bibinfo{pages}{1355} (\bibinfo{year}{2008}).

\bibitem[{\citenamefont{Altland and Zirnbauer}(1997)}]{AltlandZirnbauer1997}
\bibinfo{author}{\bibfnamefont{A.}~\bibnamefont{Altland}} \bibnamefont{and}
  \bibinfo{author}{\bibfnamefont{M.~R.} \bibnamefont{Zirnbauer}},
  \bibinfo{journal}{Phys. Rev. B} \textbf{\bibinfo{volume}{55}},
  \bibinfo{pages}{1142} (\bibinfo{year}{1997}).

\bibitem[{\citenamefont{Dyson}(1962)}]{Dyson1962}
\bibinfo{author}{\bibfnamefont{F.~J.} \bibnamefont{Dyson}},
  \bibinfo{journal}{Journal of Mathematical Physics}
  \textbf{\bibinfo{volume}{3}} (\bibinfo{year}{1962}).

\bibitem[{\citenamefont{Pruisken}(1985)}]{Pruisken1985}
\bibinfo{author}{\bibfnamefont{A.~M.~M.} \bibnamefont{Pruisken}},
  \bibinfo{journal}{Phys. Rev. B} \textbf{\bibinfo{volume}{32}},
  \bibinfo{pages}{2636} (\bibinfo{year}{1985}).

\bibitem[{\citenamefont{Ludwig et~al.}(1994)\citenamefont{Ludwig, Fisher,
  Shankar, and Grinstein}}]{Ludwig1994}
\bibinfo{author}{\bibfnamefont{A.~W.~W.} \bibnamefont{Ludwig}},
  \bibinfo{author}{\bibfnamefont{M.~P.~A.} \bibnamefont{Fisher}},
  \bibinfo{author}{\bibfnamefont{R.}~\bibnamefont{Shankar}}, \bibnamefont{and}
  \bibinfo{author}{\bibfnamefont{G.}~\bibnamefont{Grinstein}},
  \bibinfo{journal}{Phys. Rev. B} \textbf{\bibinfo{volume}{50}},
  \bibinfo{pages}{7526} (\bibinfo{year}{1994}).

\bibitem[{\citenamefont{Huckestein et~al.}(1992)\citenamefont{Huckestein,
  Kramer, and Schweitzer}}]{Huckestein1992}
\bibinfo{author}{\bibfnamefont{B.}~\bibnamefont{Huckestein}},
  \bibinfo{author}{\bibfnamefont{B.}~\bibnamefont{Kramer}}, \bibnamefont{and}
  \bibinfo{author}{\bibfnamefont{L.}~\bibnamefont{Schweitzer}},
  \bibinfo{journal}{Surface Science} \textbf{\bibinfo{volume}{263}},
  \bibinfo{pages}{125 } (\bibinfo{year}{1992}).

\bibitem[{\citenamefont{Chalker and Coddington}(1988)}]{Chalker1988}
\bibinfo{author}{\bibfnamefont{J.}~\bibnamefont{Chalker}} \bibnamefont{and}
  \bibinfo{author}{\bibfnamefont{P.}~\bibnamefont{Coddington}},
  \bibinfo{journal}{Journal of Physics C: Solid State Physics}
  \textbf{\bibinfo{volume}{21}}, \bibinfo{pages}{2665} (\bibinfo{year}{1988}).

\bibitem[{\citenamefont{Khemani and Sondhi}(2015)}]{SondhiUnpublished}
\bibinfo{author}{\bibfnamefont{V.}~\bibnamefont{Khemani}} \bibnamefont{and}
  \bibinfo{author}{\bibfnamefont{S.}~\bibnamefont{Sondhi}},
  \bibinfo{journal}{unpublished}  (\bibinfo{year}{2015}).

\bibitem[{\citenamefont{Ponte et~al.}(2014)\citenamefont{Ponte, Chandran,
  Papi\'c, and Abanin}}]{Ponte2014}
\bibinfo{author}{\bibfnamefont{P.}~\bibnamefont{Ponte}},
  \bibinfo{author}{\bibfnamefont{A.}~\bibnamefont{Chandran}},
  \bibinfo{author}{\bibfnamefont{Z.}~\bibnamefont{Papi\'c}}, \bibnamefont{and}
  \bibinfo{author}{\bibfnamefont{D.~A.} \bibnamefont{Abanin}},
  \bibinfo{journal}{Annals of Physics} \textbf{\bibinfo{volume}{353}},
  \bibinfo{pages}{196} (\bibinfo{year}{2014}).

\bibitem[{\citenamefont{Ponte et~al.}(2015)\citenamefont{Ponte, Papic,
  Huveneers, and Abanin}}]{Ponte2014b}
\bibinfo{author}{\bibfnamefont{P.}~\bibnamefont{Ponte}},
  \bibinfo{author}{\bibfnamefont{Z.}~\bibnamefont{Papic}},
  \bibinfo{author}{\bibfnamefont{F.}~\bibnamefont{Huveneers}},
  \bibnamefont{and} \bibinfo{author}{\bibfnamefont{D.~A.}
  \bibnamefont{Abanin}}, \bibinfo{journal}{Phys. Rev. Lett.}
  \textbf{\bibinfo{volume}{114}}, \bibinfo{pages}{140401}
  (\bibinfo{year}{2015}).

\bibitem[{\citenamefont{Lazarides et~al.}(2014)\citenamefont{Lazarides, Das,
  and Moessner}}]{LazaridesDasMoessner2014}
\bibinfo{author}{\bibfnamefont{A.}~\bibnamefont{Lazarides}},
  \bibinfo{author}{\bibfnamefont{A.}~\bibnamefont{Das}}, \bibnamefont{and}
  \bibinfo{author}{\bibfnamefont{R.}~\bibnamefont{Moessner}},
  \bibinfo{journal}{Phys. Rev. E} \textbf{\bibinfo{volume}{90}},
  \bibinfo{pages}{012110} (\bibinfo{year}{2014}).

\end{thebibliography}
\end{document}